\documentclass{egpubl}
\usepackage{egsr2026}
 
\SpecialIssueSubmission    %
\copyrightTextTitPag{To appear in \textit{Computer Graphics Forum} (Proc.\ EGSR 2026)}
\copyrightTextRunPag{To appear in \textit{Computer Graphics Forum} (Proc.\ EGSR 2026)}

\CGFccby

\usepackage[T1]{fontenc}
\usepackage{dfadobe}  

\usepackage{cite}  %
\BibtexOrBiblatex
\electronicVersion
\PrintedOrElectronic
\usepackage{graphicx}
\ifpdf \pdfcompresslevel=9 \fi

\usepackage{egweblnk} 

\usepackage{amsmath}
\usepackage{amssymb}
\usepackage{algorithm}
\usepackage[noend]{algpseudocode}
\DeclareMathSizes{6.5}{6}{5}{5}
\usepackage{booktabs} %
\usepackage{cleveref}
\usepackage{tikz}
\usepackage{graphicx}
\usepackage{subcaption}
\usepackage{multirow}
\usepackage[table]{xcolor}

\makeatletter
\g@addto@macro\UrlBreaks{\do\-\do\0\do\1\do\2\do\3\do\4\do\5\do\6\do\7\do\8\do\9%
  \do\a\do\b\do\c\do\d\do\e\do\f}
\makeatother

\usepackage{enumitem}
\usepackage{pdfpages} %

\title[Neural Render Proxies for Interactive and Differentiable Lighting]%
      {Neural Render Proxies for Interactive and Differentiable Lighting}

\author[S. Sancho et al.]
{\parbox{\textwidth}{\centering
Sergio Sancho$^{1,2}$, Alexander Rath$^{2}$, Marco Manzi$^{2}$, Pascal Chang$^{1,2}$\\ \vspace{4px}
Amit H. Bermano$^{1,2,3}$, Derek Nowrouzezahrai$^{4,5,6}$, Markus Gross$^{1,2}$, Marios Papas$^{2}$}
\\ 
{\parbox{\textwidth}{\vspace{14px}
\centering
\begin{tabular}{@{}ccc@{}}
$^1$ETH Zurich, Switzerland & $^2$DisneyResearch$|$Studios, Switzerland & $^3$Tel Aviv University, Israel\\[2pt]
$^4$McGill University, Canada & $^5$Mila -- Quebec AI Institute & $^6$CIFAR AI Chair
\end{tabular}
}}}

\usepackage{makecell}
\usepackage{placeins}
\usepackage{microtype}

\newcommand{\ourmethod}{neural render proxy}

\def\numx#1e#2{{#1}\mathrm{e}{#2}}

\newcolumntype{T}[1]{>{\raggedright\arraybackslash}p{#1}}

\newcommand{\lightpos}{\ell}

\newcommand{\px}[0]{\text{px}}

\usepackage{overpic}

\begin{document}

\teaser{
    \centering
    \includegraphics[width=1.0\linewidth]{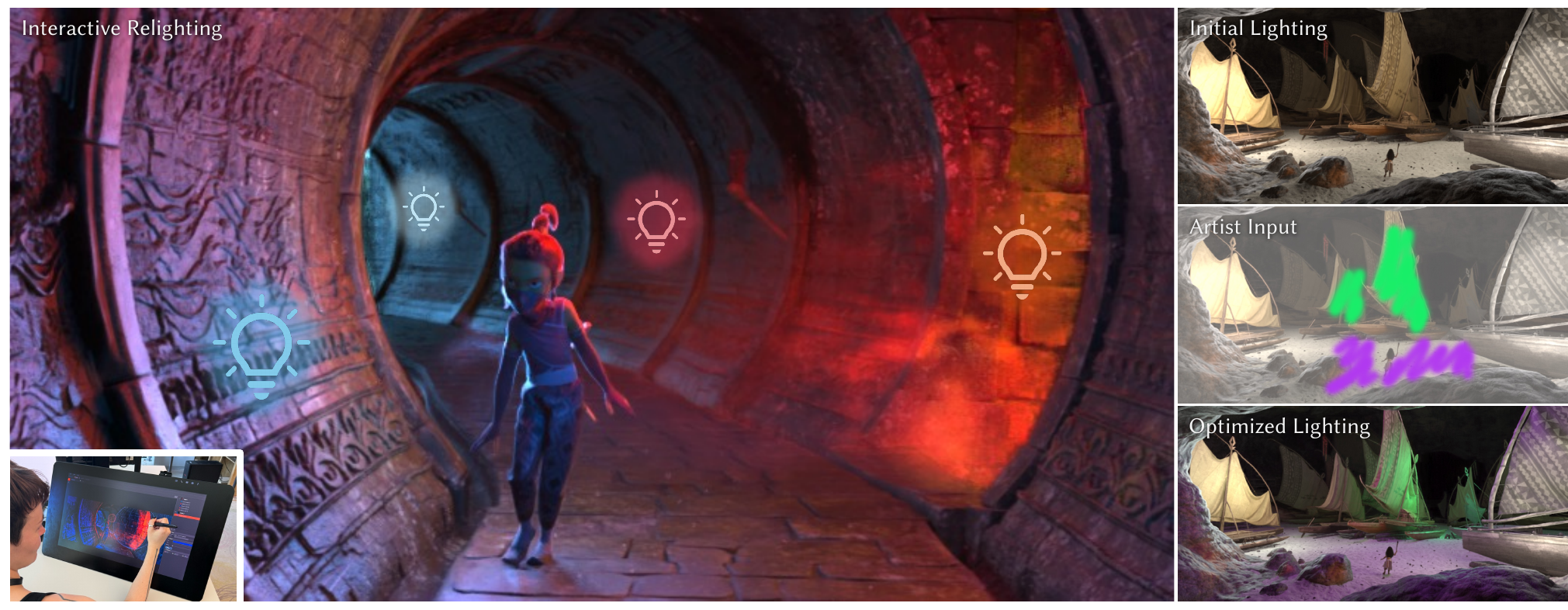}
    \caption{We train a compact network to predict transport through the scene arriving at sensor locations using light-agnostic path data, allowing us to relight complex scenes. Our model is interactive, has a modest memory footprint, generates high-quality and temporally coherent shading, and is differentiable, admitting image-driven lighting parameter optimization through otherwise non-differentiable production renderers.\\  \hfill\copyright~Disney
    }
    \label{fig:teaser}
    \vspace{10pt}
}

\maketitle

\begin{abstract}
   Within the CG animation production pipeline, the challenges artists tackle in Lighting can be immense. Even minor adjustments require re-rendering massive scenes with slow offline renderers; global illumination has to be sampled and complex shaders have to be evaluated, leading to iteration times of minutes to hours per frame.
   To accelerate this process, we introduce a novel neural render proxy (NRP) that enables differentiable relighting of static scenes with fixed camera and materials at interactive rates. Our main insight is the decoupling of traditional rendering into path sampling and emission computation. From a single, light-agnostic render pass, we collect light transport data in the form of path samples. This enables rapidly sampling lighting conditions on the fly, and training a scene-specific lightweight neural network that learns how light is transported from any scene location to any image pixel.
   This approach is compatible with non-differentiable production renderers, induces minimal memory requirements during inference, and scales only with resolution and the number of light sources and parameters, but independently of scene and appearance complexity. Our evaluation demonstrates interactive frame rates for relighting ($\sim$30--60 Hz) while closely approximating the visual fidelity of ground-truth path tracing. %
   In addition, the differentiable NRP enables fast, gradient-based inverse workflows, allowing artists to efficiently solve for lighting parameters from intuitive image-space edits or generative targets.


\end{abstract}  

\section{Introduction}

High-fidelity lighting in production environments is inherently iterative and slow. Fine-tuning lighting typically happens at the very end of the production pipeline, i.e., once materials, camera and geometry are fixed. However, even minor edits at this stage require iteration times of minutes to hours due to frame or sequence re-renders. This bottleneck is compounded by the fact that current production pipelines are effectively one-way streets: They generate images from scene parameters, but cannot translate intuitive, image-space art direction back into the 3D scene. Since production renderers are usually non-differentiable, artists are trapped in a slow trial-and-error loop, as image-guided optimization of lighting parameters is not readily supported.

Prior work on interactive relighting has largely relied on image-space post-processing (e.g., compositing), generative image synthesis methods, or interactive renderers. While compositing is an essential tool in production pipelines, it is fundamentally an image-space operation, limiting the light edits it supports. Generative AI attempts to bridge this gap in image space, but, while visually impressive, struggles to faithfully represent physical interactions and lacks controllability, limiting its use for lighting. Conversely, while modern production relies on physically accurate path tracing~\cite{TOG-prod-rendering}, interactive path tracing systems do not scale yet to the geometric and material complexity of production scenes. Consequently, digital lighting artists have no viable way to reconcile fast iteration with the requirement for physically accurate, production-scale rendering.

We introduce \ourmethod{} (NRP), an approach to interactive lighting for static scenes with fixed views that bridges the gap between high-fidelity path tracing and interactive scene- and image-space light manipulation. Our key insight is that decoupling the expensive light transport simulation from final pixel synthesis allows us to cache complex physical interactions into a highly compressed, relightable representation. By acting directly on path vertices and throughput weights generated by an offline path tracer, we train a compact multi-layer perceptron (MLP) that maps parameters of standard light sources, like sphere lights or quad lights, to their effect in the image plane. This lightweight representation enables interactive re-rendering with dynamic lighting configurations at runtime, while faithfully synthesizing complex light transport and scaling independently of scene geometry and material complexity.

To attain interactive relighting without exorbitant pre-computation, our training scheme extracts rich supervision data from a single, light-agnostic rendering pass. This distinguishes our approach from prior image-based neural transport methods, which explicitly require the generation of many different lighting configurations in an expensive pre-pass~\cite{zhang_neural_2021}. During training, we reuse our light-agnostic sample data to create training examples on-the-fly, massively reducing end-to-end turnaround times and memory footprint. Our efficient training scheme takes between tens of minutes and a few hours depending on scene complexity, allowing it to seamlessly integrate into standard overnight render cycles to be ready for interactive editing the next morning. Once trained, NRP compresses tens of gigabytes of raw path data into a few megabytes of network weights, which can be quickly loaded into memory and used for lighting at interactive frame rates (i.e., $\sim$30--60 Hz), while maintaining temporal consistency across continuous edits.

In addition, our resulting neural model is differentiable and allows us to back-propagate image-space losses to light parameters at interactive rates, enabling image-based light editing paradigms, including ``light painting'' and art direction guided by modern generative image synthesis tools.

In summary, our contributions are:
\begin{itemize}
    \item An efficient framework for re-using paths from path tracers for interactive relighting at production-quality accuracy.
    \item A neural render proxy that compresses the samples by several orders of magnitude and enables inverse rendering applications.
    \item A suite of tools based on our neural render proxy that demonstrate its applicability for art-directed and generative lighting.
\end{itemize}

\section{Related Work}

\begin{figure*}[t]
  \centering
   \includegraphics[width=1.0\linewidth]{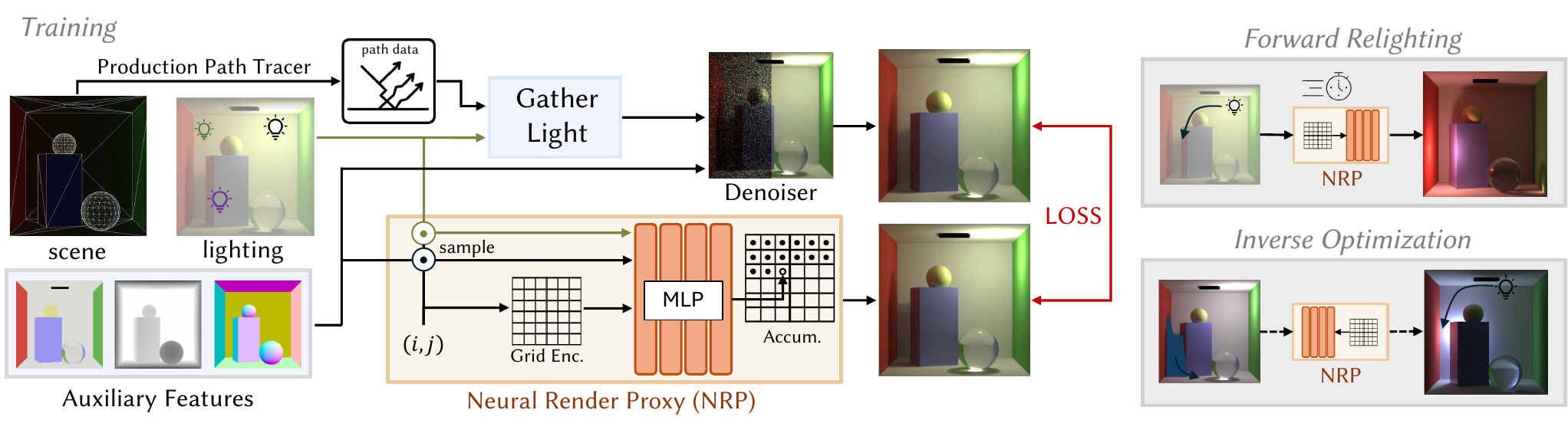}
   \vspace{-15pt}
   \caption{\textbf{Our Pipeline}. Training (left): We collect lighting agnostic paths from path tracing the scene before generating denoised image with random lighting conditions to serve as training data. Our \ourmethod{} (NRP; middle) is a lightweight MLP with hashgrid positional encoding that predicts light transport to a sensor pixel from an input light. To reduce training time and avoid spectral bias, we condition the NRP on auxiliary features (albedo, depth and normals) at sensor image coordinates. At inference time (top right), we can relight the scene at interactive rates by inserting virtual light sources into the scene. Since the NRP is also differentiable, we can optimize for lighting parameters (bottom right) using gradient descent, driven by any number of image-space guidance approaches.
   }
   \label{fig:pipeline}
\end{figure*}

\paragraph*{Neural image-based methods.} Recent neural relighting methods leverage pretrained diffusion models for scene generalization, with intrinsic decompositions into albedo, normal, and shading or material property buffers~\cite{zeng_rgbx_2024, luo_intrinsicdiffusion_2024, lyu_intrinsicedit_2025} enabling finer relighting control.
User control has been extended to text-guided lighting~\cite{zhang_scaling_2025}, scribble-based guidance~\cite{choi_scribblelight_2025}, parametric light manipulation~\cite{magar_lightlab_2025}, and intrinsics-based relighting~\cite{xing_luminet_2025}. Others condition on radiometric cues such as radiance hints~\cite{zeng_dilightnet_2024}, HDR environment maps~\cite{erel_practilight_2025, jin_neural_2024}, or point lights~\cite{bharadwaj_genlit_2025}.
Closer to our setting, LightFormer~\cite{ren_lightformer_2024} encodes virtual point lights into neural embeddings for real-time relighting of fully dynamic scenes, and RenderFormer~\cite{zeng_renderformer_2025} uses transformers to render triangle meshes with global illumination without per-scene training, though it is limited to 4,096 triangles and a single BRDF model. Neither method supports volumetric transport, and neither has been applied to inverse lighting optimization.

Earlier learning-based methods train on large datasets of images rendered under many lighting configurations~\cite{ren_image_2015, zhang_neural_2021}; more recent work improves generalization via self-supervised inverse rendering~\cite{careaga_physically_2025} or shadow-aware single-image relighting~\cite{griffiths_outcast_2022}. Despite their flexibility, most of these approaches rely on generative priors that do not adhere to production art direction and offer limited guarantees on the physical consistency of complex transport effects (caustics, inter-reflections), and are incompatible with the physics-based light transport parametrizations of production rendering workflows.

\paragraph*{Interactive and accelerated rendering.}
Production path tracers require minutes to hours per frame, making interactive lighting iteration infeasible.
Real-time denoising and upscaling techniques reduce noise in low-sample renders but do not change the fundamental cost of re-tracing paths when lights move.
Spatiotemporal reservoir resampling~\cite{bitterli_restir_2020} improves light sampling efficiency for direct illumination but still requires the full scene and renderer in the loop for each configuration.
Neural Radiance Caching (NRC)~\cite{muller_nrc_2021} trains a small MLP online to predict radiance at path vertices, allowing early path termination and accelerating convergence within a running path tracer. Though conceptually related, NRC encodes the current lighting implicitly in its weights rather than accepting light parameters as input, so it cannot evaluate radiance under novel lighting without continued online training and provides no gradient path to light parameters for inverse optimization.
Path guiding methods~\cite{muller_pathguiding_2017, SiggraphCourse2025pathguiding} improve sampling efficiency but likewise require full scene access and per-configuration re-tracing, and are not differentiable.
Our method decouples the renderer from inference entirely: Once the neural proxy is trained from a single light-agnostic rendering pass, the scene and renderer are no longer needed for either interactive forward relighting or differentiable inverse applications.

\paragraph*{Differentiability and physically-based rendering.} Production renderers are typically non-differentiable, ruling out gradient-based optimization of scene parameters from image-space objectives. Differentiable renderers address this gap through specialized automatic differentiation and just-in-time compilation~\cite{jakob_mitsuba_2022, jakob_drjit_2022}, which incorporate adjoint methods such as radiative backpropagation~\cite{nimier-david_radiative_2020} and path replay backpropagation~\cite{vicini_path_2021}. Visibility discontinuities are commonly treated through explicit boundary sampling~\cite{li_differentiable_2018}, path-space boundary integrals~\cite{zhang_pathspace_2020}, and warped-area reparameterization~\cite{bangaru_warped_2020, wu_wosc_2025}; these approaches require specialized boundary or auxiliary computations that add significant overhead. Lipp et al.~\cite{lipp_view-independent_2024} specialize adjoint light tracing for view-independent lighting design optimization at interactive rates, but like all differentiable rendering methods discussed above, require full scene access and ray tracing at each gradient step, inheriting the computational challenges of forward rendering on production-scale scenes.

Neural inverse rendering methods~\cite{boss_nerd_2021, zhang_physg_2021} and differentiable Gaussian surfel approaches~\cite{jiang_differentiable_2025} are implicitly differentiable but do not capture full physics-based volumetric transport. Hadadan et al.~\cite{hadadan_inverse_2023} use a neural radiometric prior grounded in the rendering equation for inverse global illumination, but their method targets scene recovery rather than interactive relighting.

We are most closely inspired by methods that fit local neural surrogates (i.e., proxies) for small-step optimization to (non-differentiable) path tracing output~\cite{fischer_zerograds_2024} or estimate higher-order gradients stochastically~\cite{wang_stochastic_2025}. During optimization, however, the renderer and scene must remain in the loop: Each step re-renders the scene (at low sample counts) to update the local surrogate before taking a gradient step through it. Our method requires neither re-loading the full scene nor re-tracing rays after training, compressing tens of gigabytes of path data into a few megabytes of network weights within a modest time budget. Variance-aware inverse rendering~\cite{yan_variance_2024} has recently been proposed to make the variance of Monte Carlo estimators itself differentiable and optimizable with respect to scene parameters. Our neural proxy avoids this problem entirely by producing noise-free, deterministic derivatives via standard backpropagation. This comes at a cost: The proxy is specific to a given scene and view, and introduces approximation bias from MLP regression, which we quantify in \autoref{sec:implementation}.

\paragraph*{Lighting estimation.} Methods that estimate lighting from images are commonly used for relighting and virtual object insertion. Early approaches rely on regressing the lighting parameters \cite{gardner_learning_2017}, and recent works use generative priors to recover high-fidelity HDR environment maps: Diffusion models can estimate scene lighting directly~\cite{liang_luxdit_2025} or via high-frequency reflective proxies~\cite{phongthawee_diffusionlight_2024, tong_spatiotemporally_2025}, while other approaches use spatiotemporal representations~\cite{shen_stglight_2025} or joint geometry-material-lighting recovery~\cite{wang_humanir-mgi_2025}. Environment map optimization has also been shown to benefit from diffusion-guided refinement \cite{liang_photorealistic_2025}. UniLight proposes a lighting representation suitable for transferring lighting conditions across images \cite{zhang_unilight_2025}. Whether parametric or map-based, these methods recover lighting from observations rather than providing interactive control over individual light parameters or inverse optimization from image-space objectives, which our approach enables.

\paragraph*{Traditional relighting.} Precomputed radiance transfer (PRT) projects light transport onto a compact basis for rapid relighting~\cite{prt_survey_2005}, with recent neural extensions improving flexibility but remaining restricted to distant environment map illumination~\cite{neural_prt_2022, raghavan_neural_2023}. Instead of pre-integrating against a fixed basis, we cache raw transport structure (path vertices and throughput weights), enabling relighting with arbitrary lights and materials not expressible in any pre-integration basis, including local lights in scenes with participating media (volumes). 

\paragraph*{Lighting design in computer cinematography.}
Interactive lighting in production has a long history predating modern path tracing.
Painting-based inverse systems trace back to Schoeneman et al.~\cite{schoeneman_painting_1993}, later generalized to nonlinear optimization over many light parameters via black-box renderer evaluation~\cite{pellacini_lighting_2007}.
Complementary interfaces let artists directly manipulate shadows as first-class primitives~\cite{pellacini_shadow_2002}, and user studies have compared direct, indirect, and painting paradigms~\cite{kerr_evaluating_2009}.
In parallel, deep-framebuffer relighting engines such as \textit{Lpics}~\cite{pellacini_lpics_2005} and \textit{Lightspeed}~\cite{ragankelley_lightspeed_2007} enabled interactive previews of production-scale RenderMan scenes by caching first-hit surface response and re-executing simplified light shaders on the GPU.
These systems were tied to the REYES era: direct illumination only, first-hit caches rather than full transport, strictly forward, and a bake invalidated by any change to camera, geometry, or shaders.
Modern productions have largely moved to fully path-traced pipelines~\cite{TOG-prod-rendering} in part to escape this baking overhead, at the cost of re-rendering per edit.
Our method revisits this spirit for the path-traced era: a single light-agnostic pass yields a compact, differentiable proxy that captures multi-bounce transport and supports both interactive forward relighting and gradient-based inverse design.

In summary, the methods above address complementary aspects of lighting design but leave fundamental gaps. In production, artists craft illumination by directly manipulating light parameters such as positions, intensities, and shapes, retaining full creative control at the cost of iteration speed. Research on accelerated rendering, efficient sampling, and denoising continues to narrow this gap and is indispensable, but the complexity of production scenes, with massive geometric complexity, complex global illumination, and intricate shading graphs, keeps physically grounded interactive editing out of reach. Generative and image-based methods offer intuitive, high-level control through prompts, scribbles, or reference images, but without grounding in the physical transport of the scene. The outputs of generative methods do not decompose into parametric light descriptions and cannot guarantee that a proposed edit is physically realizable.

Our method bridges both: A single offline precomputation step yields a compact, differentiable neural proxy that provides interactive forward relighting with full parametric control, while its differentiability enables gradient-based translation of image-space objectives into scene lighting parameters.

\algrenewtext{If}[1]{\textbf{if} #1:}
\algrenewtext{For}[1]{\textbf{for} #1:}
\algrenewtext{While}[1]{\textbf{while} #1:}
\algrenewtext{Function}[2]{\textbf{def} \textsc{#1}(#2):}
\renewcommand{\algorithmicindent}{0.8em}
\definecolor{passA}{HTML}{0072B2}
\definecolor{passB}{HTML}{D55E00}
\definecolor{frameGray}{HTML}{BFBFBF}
\newcommand{\algobox}[2]{%
  \begin{tikzpicture}
    \node[draw=frameGray, line width=0.4pt, rounded corners=2.5pt,
          fill=#1, inner xsep=2pt, inner ysep=4pt]{#2};
  \end{tikzpicture}}
\newcommand{\algoBodyRender}{%
  {\fontsize{6.5pt}{7.8pt}\selectfont
  \renewcommand{\algorithmicindent}{0.5em}
  \begin{algorithmic}
    \Function{Render}{px, scene, virtLights}
        \color{passB} \State $I_\px \gets \text{RGB}(0, 0, 0)$
        \For{$i = 0 \dots N_\text{spp}$}
        \color{passA} \State $T_0 \gets \text{RGB}(1, 1, 1)$
        \color{passA} \State $x_0, \omega \gets \text{SampleCamera}(\text{pixel})$
        \color{black} \For{$j = 0 \dots \text{maxDepth}$}
            \color{passA} \State $x_{j+1} \gets \text{Intersect}(\text{scene}, x_j, \omega)$
            \color{passB} \For{light $\in$ virtLights}
                \If{hit = Intersect(light, $x_j, x_{j+1}$)}
                    \State $I_\px \gets I_\px + T_j \cdot L_e(\text{light}, \text{hit})$
                \EndIf
            \EndFor
            \color{passA} \State $\omega, f_s, p \gets \text{SampleBsdf}(x_{j+1}, -\omega)$
            \State $T_{j+1} \gets T_j \cdot f_s / p$
        \EndFor
        \EndFor
    \color{passB} \State \Return $I_\px$
    \EndFunction
  \end{algorithmic}}}
\newcommand{\algoBodyDecoupled}{%
  {\fontsize{6.5pt}{7.8pt}\selectfont
  \renewcommand{\algorithmicindent}{0.5em}
  \begin{algorithmic}
    \color{passA}
    \Function{SamplePaths}{px, scene}
        \State $\mathcal{P}_\px \gets [\;]$
        \For{$i = 0 \dots N_\text{spp}$}
        \State $T_0 \gets \text{RGB}(1, 1, 1)$
        \State $x_0, \omega \gets \text{SampleCamera}(\text{pixel})$
        \For{$j = 0 \dots \text{maxDepth}$}
            \State $x_{j+1} \gets \text{Intersect}(\text{scene}, x_j, \omega)$
            \State $\omega, f_s, p \gets \text{SampleBsdf}(x_{j+1}, -\omega)$
            \State $T_{j+1} \gets T_j \cdot f_s / p$
        \EndFor
        \State $\mathcal{P}_\px.\text{append}(\,[\, (T_0, x_0), (T_1, x_1), \dots \,]\,)$
        \EndFor
    \State \Return $\mathcal{P}_\px$
    \EndFunction
  \end{algorithmic}%
  \vspace{-1.5pt}\par\noindent\textcolor{black!15}{\rule{\linewidth}{0.3pt}}\par\vspace{2pt}%
  \begin{algorithmic}
    \color{passB}
    \Function{GatherLight}{virtLights, \textcolor{passA}{$\mathcal{P}_\px$}}
        \State $I_\px \gets \text{RGB}(0, 0, 0)$
        \For{$[ \, (T_0, x_0), (T_1, x_1), \dots \, ] \in \mathcal{P}_\px$}
        \For{$j = 0 \dots \text{maxDepth}$}
            \For{light $\in$ virtLights}
                \If{hit = Intersect(light, $x_j, x_{j+1}$)}
                    \State $I_\px \gets I_\px + T_j \cdot L_e(\text{light}, \text{hit})$
                \EndIf
            \EndFor
        \EndFor
        \EndFor
    \State \Return $I_\px$
    \EndFunction
  \end{algorithmic}}}
\newsavebox{\sbAlgoD}
\newlength{\algoContentW}
\newlength{\algoBoxHt}
\newlength{\algoHtTmp}
\newcommand{\algoframeSnug}[1]{%
  \algobox{gray!4}{\begin{minipage}{\algoContentW}#1\end{minipage}}}
\newcommand{\algoCell}[1]{%
  \begin{minipage}[b][\algoBoxHt][c]{\linewidth}\centering\algoframeSnug{#1}\end{minipage}}
\begin{figure}[!th]
    \centering
    \setlength{\algoContentW}{\dimexpr0.49\linewidth-6pt\relax}%
    \savebox{\sbAlgoD}{\algoframeSnug{\algoBodyDecoupled}}%
    \settoheight{\algoBoxHt}{\usebox{\sbAlgoD}}%
    \settodepth{\algoHtTmp}{\usebox{\sbAlgoD}}%
    \addtolength{\algoBoxHt}{\algoHtTmp}%

    \begin{subfigure}[b]{0.49 \linewidth}
        \centering
        \algoCell{\algoBodyRender}
        \caption{Standard path tracing interleaves \textcolor{passA}{tracing rays} and \textcolor{passB}{illumination computation}, requiring information about scene and lights at the same time.}
        \label{alg:classic_pt}
    \end{subfigure} \hfill
    \begin{subfigure}[b]{0.49 \linewidth}
        \centering
        \algoCell{\algoBodyDecoupled}
        \caption{We decouple \textcolor{passA}{tracing rays} (only requiring scene information) and \textcolor{passB}{producing the image} (based on light parameters and pre-computed paths).}
        \label{alg:decoupled}
    \end{subfigure}
    
    \caption{Classic path tracing vs. our decoupled approach}
    \label{fig:both_algorithms}
\end{figure}

\begin{figure*}[t]
  \centering
   \includegraphics[width=1\linewidth]{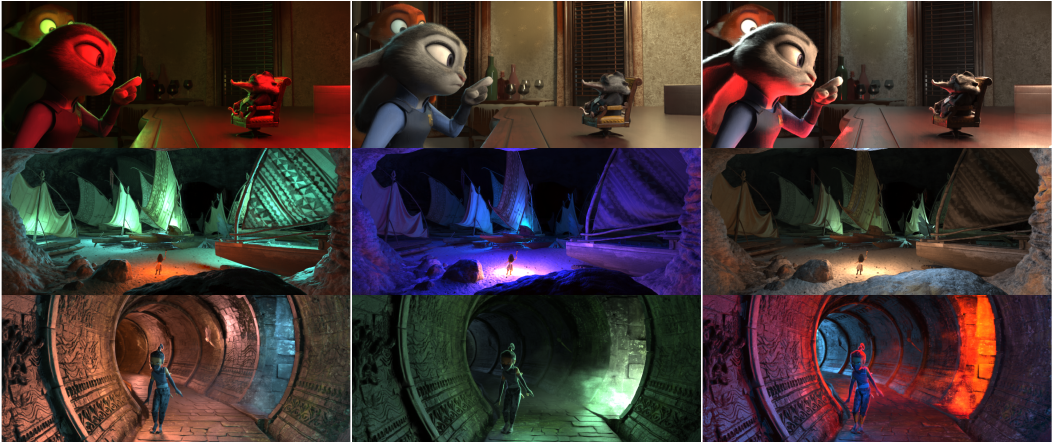}
   \caption{\textbf{Production scenes.} Interactive relighting results created by artists using our \ourmethod~on large-scale production scenes. Each row shows three different light configurations for the same scene. Our method faithfully captures complex light transport effects including fur (top), translucency (middle), and volumetric lighting (bottom). \copyright~Disney
   }
   \label{fig:results_production}
\end{figure*}

\section{Method}
\label{sec:method}
The traditional offline rendering pipelines for relighting of static scenes suffer from slow scene loading, long render times, and are typically non-differentiable. We replace this with a differentiable neural render proxy trained from data collected during a single rendering process that can predict images for arbitrary lighting configurations at interactive rates and is amenable to inverse rendering applications.

We begin by showing that in the context of forward path tracing and virtual lights, 
rendering can be decoupled into path sampling, which requires scene but not light information, and emission computation, which requires light but no scene information (\autoref{sec:decoupled_rendering}).
This decoupling allows us to pre-compute path samples once, and reuse them for fast relighting with arbitrary light configurations, albeit at the cost of operating on large path sample dumps.

To reduce memory requirements and support differentiability, we introduce our \ourmethod~(\autoref{sec:neural_render_proxy}), which encodes the paths' sample information in a compressed, continuous, and differentiable representation, enabling applications such as art-directed edits and generative relighting.

\subsection{Decoupled Rendering}
\label{sec:decoupled_rendering}
Virtual light sources are popular in production rendering because they allow adding emission to a scene without introducing undesired scattering effects---when a virtual light is intersected, its emission is added to the image, but the ray continues unaltered.
Since the construction of paths is therefore unaffected by the placement of light sources, we can re-order the computations carried out by a path tracer without altering the result.
Traditionally, path tracing interleaves the intersection of virtual lights and the computation of their emission with the path generation process (\autoref{alg:classic_pt}).
In our approach, we split the computation into two passes (\autoref{alg:decoupled}):

\paragraph*{The first pass}(\textsc{SamplePaths}) samples paths and accumulates path vertex positions and their corresponding throughput weights for a given originating pixel.
This pass absorbs the most expensive components of the rendering process: the cost of loading the scene, intersecting rays with its geometry, and evaluating shaders.
Since it does not require any information about the light sources, the aggregated data can be reused to render various different lighting configurations.

\paragraph*{The second pass}(\textsc{GatherLight}) uses the path samples to test each segment for intersection with virtual lights.
When a light is intersected, its emission is weighted by the throughput weight at the current vertex and simply added to the image.
Utilizing the high memory bandwidths of GPUs, this relighting process can be carried out at interactive rates without requiring advanced acceleration structures.
However, this pass requires storing large sample dumps alongside each render, typically weighing dozens of gigabytes per frame, which can quickly become prohibitive, e.g., when relighting of entire animation sequences is desired.

\paragraph*{Importance sampling.}
To ensure that \textsc{GatherLight} can produce unbiased renders for novel lighting configurations, \textsc{SamplePaths} needs to sample all possible directions that could intersect a light.
Fortunately, illumination-unaware sampling techniques such as BSDF sampling fulfill this requirement.
Similarly, techniques such as Russian roulette can be employed to reduce the number of path samples, and hence the computational cost of \textsc{GatherLight}.
However, our method does not benefit from illumination-aware techniques such as next event estimation (NEE) or path guiding, as these require knowledge of the light configuration at path generation time.
In practice, the noise of the generated image depends on the number of segments that intersect with the virtual light. This means that very small light sources as well as lights located in sparsely sampled scene regions can be challenging for our approach. %
We discuss our mitigation for this issue further in \autoref{sec:denoising}.

\paragraph*{Volume rendering.}
Our method naturally extends to volumes that can be rendered through analogous free-flight sampling techniques, including both homogeneous and heterogeneous volumes with arbitrary phase functions, but excludes volumes with spectrally varying densities.
Free-flight sampling techniques, such as delta tracking or analog decomposition tracking \cite{kutz2017spectral}, distribute scattering distances according to transmittance, which cancels out changes to the throughput weight that would otherwise be required when intersecting an object within the volume.
Since such methods do not require light source locations, and all information about transmittance is fully contained in the distribution of path segments,  \textsc{GatherLight} naturally extends to volumetric effects rendering without any alterations.

\begin{figure*}[t]
  \centering
   \includegraphics[width=\textwidth]{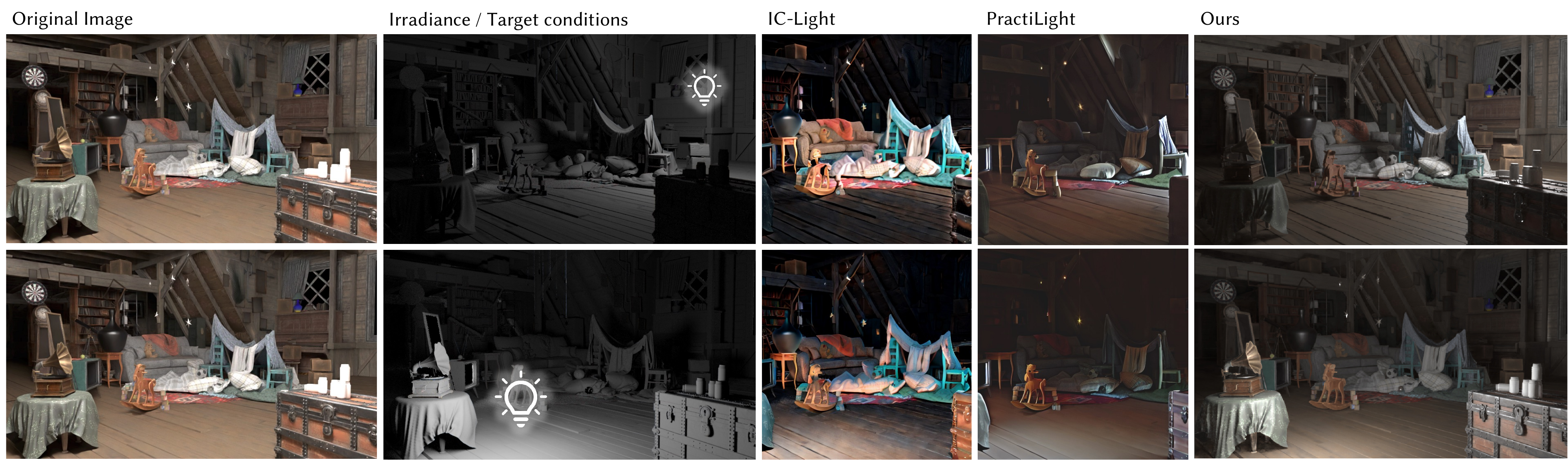}
   \caption{\textbf{Qualitative image-based relighting comparison.} Our relighting setting matches that of image-based relighting using generative networks. Two leading approaches with similar input settings (image, auxiliary features, and lighting condition) are presented. Generative models do not preserve scene geometry, do not adhere to material appearance behavior, and produce lighting that is plausible, but is not physically reasonable. Our method produces high-quality and high-fidelity appearance in real-time. Note that the aspect ratio is different since the generative models used consider squared images.}
   \label{fig:comparison_diffusion}
\end{figure*}

\subsection{Differentiable Neural Render Proxy}
\label{sec:neural_render_proxy}
To reduce the storage requirements of our relighting approach, we train a neural network to imitate the \textsc{GatherLight} function.
Since this function has dynamic inputs, i.e., the virtual lights and their parameters, whose count and role depend on the number of light sources and their respective types, we begin by decomposing the function.
By utilizing the linearity of light transport, we can split the general \textsc{GatherLight} function into distinct $\textsc{Gather}_\text{type}$ functions for individual light source types (e.g., quad lights, sphere lights) and factor out the intensity of the light source $E(v)$,
\begin{equation}
    \textsc{GatherLight}(\mathcal{V}, \mathcal{P}_\px) = \sum_{v \in \mathcal{V}} E(v) \cdot \textsc{Gather}_{\text{type}(v)}(v, \mathcal{P}_\px).
\end{equation}
Here, $\mathcal{V}$ is the list of all virtual light sources, with each entry $v~\in~\mathcal{V}$ containing all the parameters relevant for the respective light type.
$\mathcal{P}_\px$ is the list of all paths for a given pixel $\px$ as traced by $\textsc{SamplePaths}$, where each entry $p = [ \, (T_0, x_0), (T_1, x_1), \dots \, ] \in \mathcal{P}_\px$ corresponds to one path generated by path tracing, with $x_i$ being the path vertex at index $i$, and $T_i$ its corresponding throughput weight. We denote the computed pixel value from the path sample data by $I_{\text{px}}=\textsc{GatherLight}(\mathcal{V}, \mathcal{P}_\px)$, and refer to it as the \emph{reconstruction} throughout the paper.

We then train one neural network for each type of light source on the pre-computed path sample dump to obtain
\begin{equation}
\label{eq:network-output}
    \mathcal{N}_\text{type}(\text{px}, F_\px, v) \approx \textsc{Gather}_{\text{type}}(v, \mathcal{P}_\px) .
\end{equation}
Inspired by denoising architectures, we have found that including auxiliary pixel features $F_\px$, such as the albedo or normals at a given pixel $\px$, drastically improves the accuracy of the network and accelerates its training convergence.
The exact features and encodings used by our model are discussed in \autoref{sec:network_architecture}.

Computing the final image given a set of light sources $\mathcal{V}$ is then achieved by summing up the result of our neural network for each light source and weighting by the emissive strengths $E$ of the lights:
\begin{equation}
\label{eq:neural_image}
    \hat{I}_\text{px} = \sum_{v \in \mathcal{V}} E(v) \cdot \mathcal{N}_{\text{type}(v)}(\px, F_\px, v) .
\end{equation}

\paragraph*{Scaling with parameters.}
The number of input parameters to our neural networks greatly depends on the types of light sources to be rendered.
We have observed that our method works best for lights with lower parameter counts, such as sphere lights (four parameters: 3D position, radius) or quad lights (eight parameters: 3D position, 3D normal, width and height).
Generalizing to more complex types, such as area lights of arbitrary shapes or textured lights, remains a question for further investigation. 

\paragraph*{Differentiability.}
Replacing \textsc{GatherLight} by a neural network not only drastically reduces the storage requirements (typically from gigabytes to megabytes), but also converts an otherwise non-differentiable function to a differentiable one.
This makes our model an effective surrogate for the full renderer, which we exploit for applications such as art-directed edits to light configurations (see \autoref{sec:app_art_edits}) and harnessing generative models for automatic lighting of scenes (see \autoref{sec:app_generative}).

\section{Implementation}
\label{sec:implementation}

We have implemented the \ourmethod{} described in \autoref{sec:method} on top of PyTorch and \texttt{tiny-cuda-nn} \cite{tiny-cuda-nn}.
The overall process is depicted in Figure~\ref{fig:pipeline}.%

\subsection{Path data caching}
We collect path data by rendering the scene offline once in a light-agnostic pass as described in the \textsc{SamplePaths} step in \autoref{sec:decoupled_rendering}: Paths are traced from the camera following BSDF importance sampling without NEE, applying throughput-based Russian roulette. For production scenes we use Walt Disney Animation Studios’ CPU production renderer Hyperion \cite{hyperion}, while academic scenes are rendered with Mitsuba 3~\cite{jakob_mitsuba_2022}. For each path vertex, we store its 3D position and the throughput weight, along with the direction for paths that escape the scene without hitting geometry. Importantly, this caching requires no intrusive or algorithmic modifications to a standard Monte Carlo path tracer. In production renderers, existing path recording structures such as those used for path guiding~\cite{SiggraphCourse2025pathguiding} can be repurposed directly. In our experiments with production scenes at a resolution of 960$\times$402 with 512 samples per pixel (spp), rendering the path data took $\sim$30 minutes including scene loading, using 32 threads of an Intel Xeon Gold 6342 at 2.8 GHz. The memory footprint of these scenes during rendering was between 14 and 28 GB of RAM. The complete details are included in the appendix.

\subsection{GPU-accelerated reconstruction}
The \textsc{GatherLight} step described in \autoref{sec:decoupled_rendering} requires testing millions of path segments for intersection, which can severely bottleneck performance if implemented naively. We thus implement a single fused Triton kernel \cite{tillet2019triton} that evaluates all segments uniformly in one pass.

\paragraph*{Memory layout and compression}
We pack the dataset into an interleaved representation, storing geometry in half-precision~(\texttt{fp16}) and HDR color throughput in \texttt{rgb9e5}, a 32-bit shared-exponent floating-point format that captures the full HDR range of path throughputs while saving 8 bytes per segment compared to a standard \texttt{float32} RGB triple. This layout lets the path data for a render of $960\times402$ at 512~spp or, equivalently, $1920\times804$ at 128~spp, both with up to 6 segments per path, fit comfortably in consumer GPU memory.

\subsection{Network architecture}
\label{sec:network_architecture}

Beyond the parameters of the light source, our networks accept nine further inputs: the pixel coordinates $\text{px}$ (2D) and pixel features $F_\px$ (albedo + depth + normal, 7D).
The outputs of the networks are the contribution of the light as described in \autoref{eq:network-output}, which are multiplied with the light emission $E$ and summed up to give the final pixel value $\hat{I}_\text{px}$.

\paragraph*{Encodings}
The pixel coordinates $\text{px}$ are encoded by a two-dimensional multi-resolution hashgrid \cite{muller_instant_2022}, which we have found helps the network to learn sharper discontinuities in image space (e.g., shadow edges).
We have experimented with encodings for the other inputs, but have found little benefit in them.

\begin{table}[t!]
\centering
\caption{Reconstruction by spp (top) and inference performance by network size (bottom) for the Kitchen scene ($885 \times 500$ pixels).}
\label{tab:timing_small_kitchen}
\resizebox{\columnwidth}{!}{%
\begin{tabular}{lcccccc}
\toprule
\textbf{Reconstruction} & 32 spp & 64 spp & 128 spp & 256 spp & 512 spp & Den. \\
\midrule
Time (ms) & 1.05 & 1.86 & 3.48 & 6.90 & 13.89 & 2.60 \\
GPU Memory (GB) & 1.22 & 2.42 & 4.82 & 9.63 & 19.25 & --- \\
\midrule\midrule
\textbf{Inference} & 16/4 & 32/4 & 64/4 & 128/8 & 256/8 & 512/8 \\
\midrule
Time (ms) & 0.83 & 0.98 & 1.20 & 3.13 & 9.13 & 31.21 \\
Model Size (MB) & 5.6 & 5.6 & 5.7 & 6.1 & 7.4 & 13.3 \\

\bottomrule
\end{tabular}%
}
\end{table}

\subsection{Training}
For each training batch, we sample random light parameters for each pixel $\px$ and construct the desired target pixel values via path data gathering. We then compute the loss between the predicted output of our network and the target output. With a standard configuration~(a $680\!\times\!680$-pixel scene at 512 spp and a network of 8 layers with 256 neurons), we train our neural proxy for 100k iterations, taking approximately 1 hour on an NVIDIA RTX 5090 (32 GB VRAM). Robust training, however, requires careful design of the light parameter sampling strategy, the loss function, and the mitigation of excessive noise in the training data. To address the latter, we denoise pixels using an off-the-shelf interactive denoiser, as described below.

\paragraph*{Denoising.}
\label{sec:denoising}
While it is possible to reconstruct individual pixel values $I_\px$ for any light configuration through gathering, we found the noise of the reconstructed values for small light sources to be prohibitive for effective network training.
To combat this, we incorporate an interactive denoiser \cite{OpenImageDenoise} in our training pipeline. This denoiser uses the same auxiliary features used by the network itself~(\autoref{sec:network_architecture}) to better preserve geometrical edges. However, utilizing a denoiser forces us to use the same light parameters across all pixels, which heavily limits lighting diversity in a single batch and hinders the training process.
To mitigate this, we keep a pool of denoised images for various light configurations in memory, and for each pixel sample uniformly from this pool.
In all our experiments, we use an image pool of size 300, and replace 2 images every 5 iterations of training.
This replacement strategy ensures adequate coverage of the domain of possible light configurations.

\paragraph*{Sampling light positions.}
To pick a random light position for a given pixel $\px$, we use different strategies depending on the characteristics of the scene. For smaller scenes, we uniformly pick among our recorded path segments for that pixel, and then choose a random point uniformly on that segment. This implicitly importance samples positions that have higher contribution to the image (e.g., are closer to the camera or strongly reflected by objects), and avoid positions that do not contribute to any pixel in the image (e.g., lights within closed objects).
However, some of our tested production scenes are gigantic in scale and only a tiny fraction is directly visible from the camera. In such cases, segment-based sampling can produce suboptimal training candidates, causing the network to allocate capacity to regions where lights are unlikely to be placed during inference. As a practical remedy for such scenes, we instead sample uniformly within the bounding box of all positions visible from the camera. More sophisticated sampling schemes could be explored in future work.

\paragraph*{Loss function}
We use the relative mean squared error loss of Müller et al.~\cite{muller_nrc_2021} to account for the high dynamic range of the data, assigning equal importance to errors across dark and bright regions:
\begin{equation}
    \mathcal{L}_{HDR}(\hat{I}, I) = \frac{1}{N_\px} \sum_\px \frac{(\hat{I}_\px - I_\px)^2}{sg(\hat{I}_\px)^2 + \epsilon},
\end{equation}
where $\hat{I}_\px$ is the predicted value at pixel $\px$, $I_\px$ is the denoised reconstructed value, $\epsilon$ is a small constant (we always set it to 0.01) and $sg(\cdot)$ is a stop-gradient operation. %

\section{Evaluation}

\begin{figure}[t!]
  \centering
   \includegraphics[width=1.0\columnwidth]{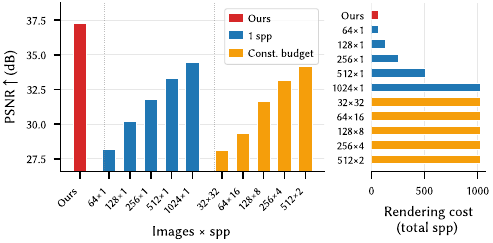}
  \caption{\textbf{Path-based training.} PSNR (dB) in tonemapped space, averaged across three scenes. Each bar is a network trained on \textit{N} images at \textit{S} spp with NEE (N$\times$S). Blue: increasing dataset size at 1 spp. Orange: fixed budget of 1024 total samples per pixel, redistributed across images. Dashed red line: our method, trained on light-agnostic path data from a 64 spp render. Despite 16$\times$ fewer sampled paths, our approach outperforms all image-based configurations.}
   \label{fig:comparison_image_based}
\end{figure}

We evaluated our system across several academic and production scenes at different sampling rates to measure scalability. All our trainings and experiments are conducted on an NVIDIA RTX 5090~(32 GB VRAM).

\subsection{Path-based Sampling}

Our method trains on path data from a light-agnostic rendering pass from which arbitrary light configurations can be evaluated on-the-fly. This allows the network to see a far greater variety of lighting conditions than any pre-rendered image dataset of comparable size.
To validate this advantage, in Figure~\ref{fig:comparison_image_based}, we compare against an image-based baseline where networks are trained on datasets of rendered images with NEE and multiple importance sampling, using light configurations sampled from the same distribution as our method.
We evaluate two settings: (1) a varying number of 1 spp images (64 to 1024), and (2) a constant rendering budget of 1024 total sampled paths per pixel, distributed across different numbers of images (e.g., 32 images at 32 spp, 512 images at 2 spp).
Note that our path-based approach uses 64 spp, so the constant-budget setting represents a 16$\times$ higher render time.

\begin{figure}[t!]
  \centering
   \includegraphics[width=1.0\columnwidth]{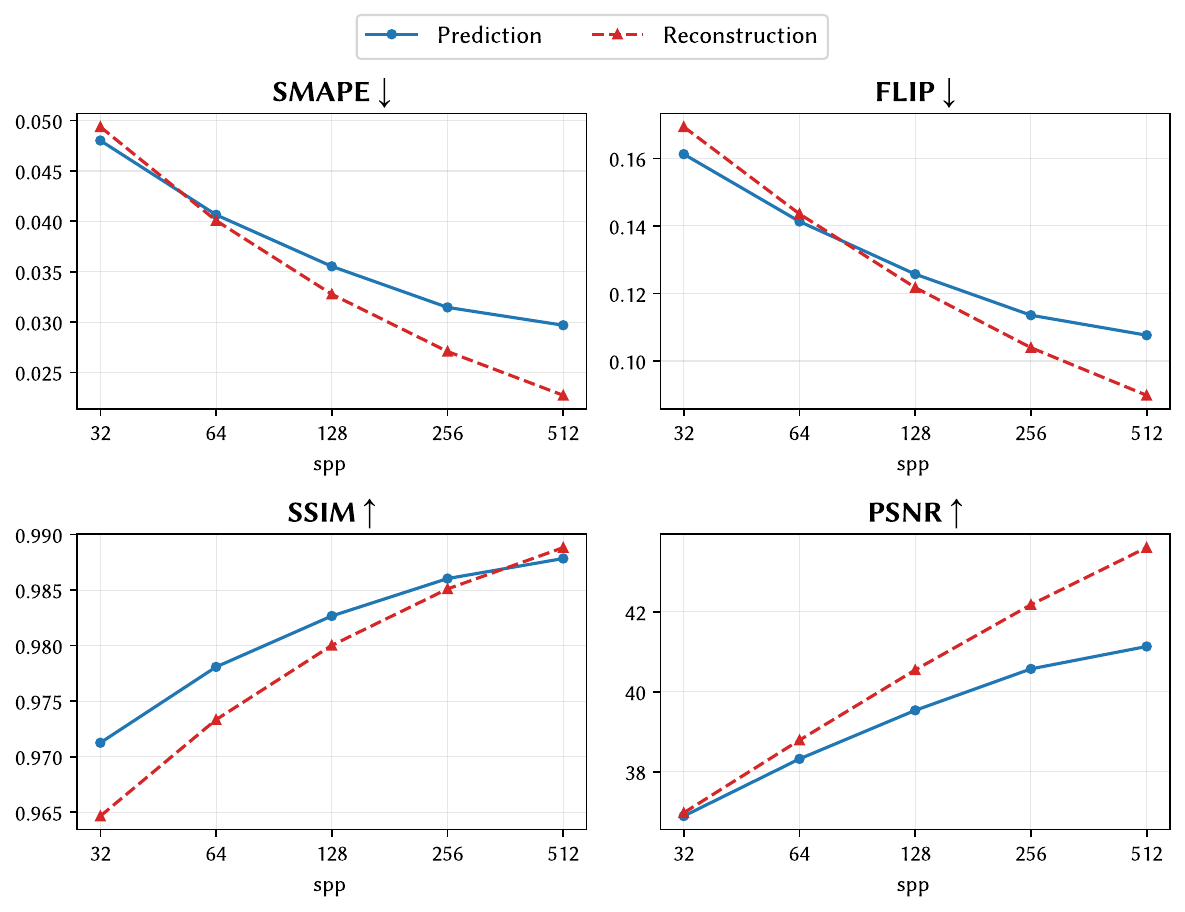}
\caption{\textbf{Quality metrics vs.\ samples per pixel.}
Mean prediction quality for network size $256\!\times\!8$ (blue) and denoised reconstruction from the \textsc{GatherLight} pass (red), averaged across three scenes.
Higher spp improves both reconstruction and prediction quality.
Notably, at low spp the learned model can outperform the reconstruction it was trained on.
Unless stated otherwise, we use 512\,spp for training.}

   \label{fig:fig_5_2_spp_ablation}
\end{figure}

We report the PSNR in tonemapped space, averaged across three scenes and evaluated on 500 validation images whose light configurations are sampled from the same distribution used during training. Our method outperforms all image-based configurations by a significant margin. Even with 1024 training images, the image-based baseline remains 2.8~dB below ours. The constant-budget ablation further shows that redistributing a larger rendering budget across fewer, higher-quality images does not close the gap. This confirms that the key advantage of our approach is not the total rendering cost, but the ability to reuse path data across many light configurations during training, providing the network with far richer supervision over the light transport distribution of the scene than a fixed dataset of images can offer.

\subsection{Training \& Network Architecture}

\paragraph*{Increasing spp for path data.}
We evaluate the effect of samples per pixel (spp) in the training path data. Figure~\ref{fig:fig_5_2_spp_ablation} shows prediction quality and the denoised reconstruction from the \textsc{GatherLight} pass across four metrics, averaged over three scenes. Increasing spp improves both reconstruction and prediction quality. Notably, at low spp (32 and 64), the network outperforms the denoised reconstruction. Since the denoiser operates independently per frame, it introduces temporal flickering across light configurations. The network, trained under many lighting conditions, learns a temporally smooth approximation of the scene's light transport. At higher spp, reconstruction noise diminishes and the denoised reference surpasses the network, whose finite capacity limits its ability to capture all high-frequency detail.

\begin{table}[t!]
\centering
\caption{Component ablation study (averaged per scene group)}
\label{tab:5_3_encoding_ablations_avg}
\resizebox{\columnwidth}{!}{
\begin{tabular}{c l c c c c}
\toprule
& Configuration & SMAPE $\downarrow$ & PSNR $\uparrow$ & SSIM $\uparrow$ & FLIP $\downarrow$ \\
\midrule
\multirow{5}{*}{\rotatebox[origin=c]{90}{\textit{Mitsuba}}} & None & 0.0533 & 35.2913 & 0.9488 & 0.1546 \\
 & Aux. & \underline{0.0290} & \underline{39.9576} & 0.9848 & \underline{0.1017} \\
 & Aux. + Den. & 0.0296 & 39.9304 & \underline{0.9851} & 0.1084 \\
 & Aux. + Enc. & 0.0566 & 34.7786 & 0.8570 & 0.1232 \\
 & Aux. + Enc. + Den. & \textbf{0.0263} & \textbf{41.0686} & \textbf{0.9885} & \textbf{0.1014} \\
\bottomrule
\end{tabular}
}
\end{table}

\begin{figure}[t!]
  \centering
   \includegraphics[width=1.0\columnwidth]{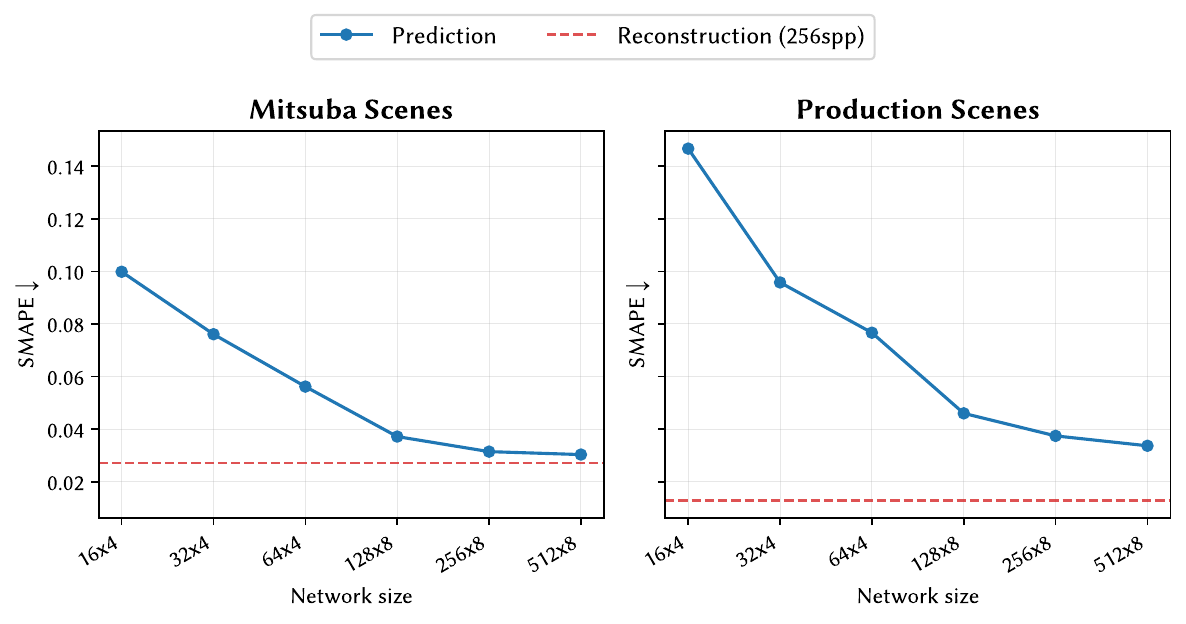}
\caption{\textbf{Effect of network size on prediction quality (SMAPE).}
Left: average over the three Mitsuba scenes; right: production scene (\emph{Cave}).
The dashed line indicates the Monte Carlo reconstruction at the same sample budget.
Larger networks consistently improve quality, with diminishing returns beyond $256\!\times\!8$.}
\label{fig:5_5_netsize_ablation_averaged}
\end{figure}

\begin{figure*}[t]
  \centering
   \includegraphics[width=1.0\linewidth]{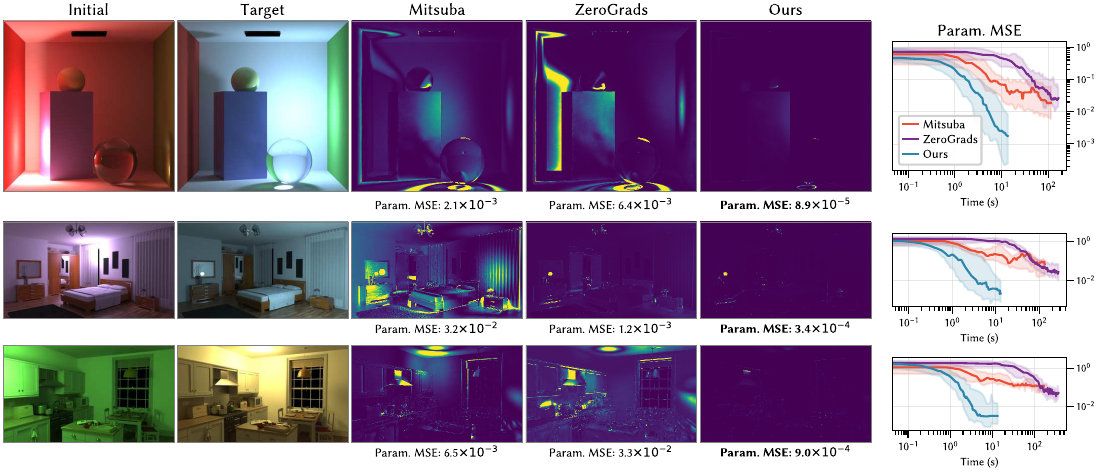}
   \vspace{-20pt}
   \caption{\textbf{Inverse optimization comparison.} A light source is optimized from a random initial position to match a target appearance. Error maps (middle) show the relative squared error of path-traced re-renders against the target. Our NRP approach converges to lower parameter error in less time than traditional (Mitsuba) and neural (ZeroGrads) baselines (right). This comparison serves as validation: Baselines operate at low spp for tractable runtimes, and their accuracy could improve with higher sample counts at proportionally longer optimization times. }
   \label{fig:5_6-comparison_differentiability}
\end{figure*}

\paragraph*{Component ablation.}

Table~\ref{tab:5_3_encoding_ablations_avg} reports the effect of three components, which is visualized in an example in the supplemental material: auxiliary features (albedo, normals, depth), denoising the reconstruction targets at training, and a hashgrid encoding for pixel coordinates. Auxiliary features substantially improve all metrics by providing geometric and material information complementary to path throughput weights. The denoiser alone does not yield significant gains, as the network already smooths noise through exposure to many light configurations. The hashgrid encoding alone degrades performance by enabling overfitting to high-frequency noise patterns. However, combining the encoding with denoised targets resolves this: Clean supervision allows the encoding to faithfully capture spatial detail without noise overfitting. The full configuration (Aux. + Enc.\ + Den.) achieves the best results across all metrics.

\paragraph*{Network size.}

\autoref{fig:5_5_netsize_ablation_averaged} shows the effect of network capacity on prediction quality, evaluated on 2k random validation images for both Mitsuba scenes (averaged over three scenes) and a production scene. Increasing the network size consistently improves quality, but with diminishing returns beyond $256\!\times\!8$ (8 layers of 256 neurons). Conversely, the bottom of \autoref{tab:timing_small_kitchen} shows that inference time keeps growing with network size, even where prediction quality has already plateaued. We therefore adopt $256\!\times\!8$ as a good compromise between inference speed, memory footprint, and prediction quality. These conclusions hold across scenes of varying complexity, including those with caustics and volumes.

\subsection{Light Optimization}
\label{sec:optimization}

We evaluate the differentiability of NRP by optimizing light parameters for sphere lights, i.e., center position $\lightpos_i \in \mathbb{R}^3$, radius $r_i \in \mathbb{R}_{>0}$, and color $E_i \in \mathbb{R}_{>0}^{3}$, to match a target image.

\paragraph*{Optimization formulation.}
For each pixel $\text{px}$, the predicted image is computed as the sum of contributions from $N$ lights:
\begin{equation}
  \hat{I}_\text{px} = \sum_{i=1}^{N} E_i \cdot \mathcal{N}_\text{sphere}(\text{px},\, \lightpos_i,\, r_i) ,
\end{equation}
where $\mathcal{N}_\text{sphere}$ is the pre-trained NRP for sphere lights.
Here, $\mathcal{N}_\text{sphere}$ denotes a trained instance of $\mathcal{N}_\text{type}$ from \autoref{eq:network-output}, specialized to sphere lights.
The loss is the mean squared error on Reinhard-tonemapped values against the target $I^*_\text{px}$:
\begin{equation}
  \mathcal{L} = \frac{1}{|\mathcal{S}|} \sum_{\px \in \mathcal{S}} \left\| \mathcal{T}\!\left(\hat{I}_\text{px}\right) - \mathcal{T}\!\left(I^*_\text{px}\right) \right\|^2, \quad \text{where} \quad
\mathcal{T}(I) = \frac{I}{1 + I},
\end{equation}
and $\mathcal{S}$ is the set of evaluated pixels (the full image, or a stochastic subset as described in \emph{Mini-batch SGD}).

Light parameters are reparameterized to unconstrained space: position and radius via the logit function $\sigma^{-1}(x) = \log(x / (1-x))$ mapping from their respective bounded domains $[\boldsymbol{\lightpos}_{\min}, \boldsymbol{\lightpos}_{\max}]$ and $[r_{\min}, r_{\max}]$, and color via the inverse softplus $\log(e^x - 1)$ mapping from $\mathbb{R}_{>0}$. Optimization is performed with Adam in unconstrained space and constrained values are recovered via the sigmoid and softplus functions at each step. For multi-light scenes, gradients are accumulated sequentially per light to bound GPU memory.

\paragraph*{Comparison against baselines.}
In order to evaluate our method in inverse rendering tasks for lighting, we compare it against Mitsuba~3~\cite{jakob_mitsuba_2022}, an academic differentiable renderer, and ZeroGrads~\cite{fischer_zerograds_2024}, a neural proxy method for black-box renderers. Note that Mitsuba~3 is a general-purpose differentiable renderer with gradients computed from the proper light transport of the scene. This evaluation does not intend to claim a superiority over real physically-based inverse renderers, but just to validate our method as an alternative when these renderers are not available, like in the context of complex production path tracers that drive our problem.

We perform 50 optimization runs per scene (Cornell Box, Bedroom, and Kitchen) with a single light and randomly sampled initial and target parameters. All methods run for 500 iterations with Adam (lr=0.05). Mitsuba uses a progressive spp schedule, starting at 8 spp for the first 100 iterations, increasing to 16 spp for iterations 100--299, and 32 spp for iterations 300--499, to reduce gradient variance as the optimization converges. ZeroGrads uses a fixed 8 spp throughout, training its neural proxy on-the-fly with a proxy learning rate of 0.001 and batch size 4. For both baselines, we use DrJIT's \texttt{@dr.freeze} decorator to eliminate JIT recompilation overhead and report only the effective computation time.

Our method does not require samples per pixel as it uses a pre-trained NRP, whereas Mitsuba~3 requires progressively increasing spp for variance reduction, and ZeroGrads requires additional hyperparameters for proxy learning. In contrast to our method, which operates on a finite set of precached paths, ZeroGrads can be run in a streaming fashion, progressively increasing accuracy at the expense of additional computation. Being non-progressive is a deliberate design choice in our approach that avoids loading bulky production scenes; incorporating progressive streams remains a promising direction for future work. Figure~\ref{fig:5_6-comparison_differentiability} shows convergence plots and error maps for representative optimization runs.

\paragraph*{Mini-batch Stochastic Gradient Descent (SGD).}
\label{sec:ablation}

\begin{table}
  \centering
  \caption{\textbf{Cache optimization ablation: Mini-batch SGD.} PSNR (dB, higher is better) is computed on exposure-normalized Reinhard-tonemapped images re-rendered with Mitsuba, averaged across three scenes. Time is for 500 optimization iterations.}
  \label{tab:cache_ablation}
  \small
  \setlength{\tabcolsep}{3pt}
  \begin{tabular}{l cc cc cc cc}
    \toprule
    & \multicolumn{2}{c}{1 Light} & \multicolumn{2}{c}{3 Lights} & \multicolumn{2}{c}{5 Lights} & \multicolumn{2}{c}{20 Lights} \\
    \cmidrule(lr){2-3} \cmidrule(lr){4-5} \cmidrule(lr){6-7} \cmidrule(lr){8-9}
    Pix.\ frac. & PSNR & Time & PSNR & Time & PSNR & Time & PSNR & Time \\
    \midrule
    \textit{Initial} & 14.3 & --- & 14.9 & --- & 15.8 & --- & 19.2 & --- \\
    \midrule
    $100\%$ & 39.5 & 21s & 29.6 & 63s & 32.1 & 138s & 32.4 & 551s \\
    $50\%$ & 39.4 & 11s & 29.9 & 33s & 32.0 & 55s & 32.6 & 286s \\
    $25\%$ & 39.4 & 5.7s & 29.9 & 17s & 32.2 & 28s & 32.5 & 149s \\
    $5\%$ & 39.1 & 1.7s & 29.8 & 4.8s & 32.4 & 7.9s & 32.6 & 31s \\
    $1\%$ & 39.0 & 0.9s & 30.7 & 2.4s & 32.0 & 3.9s & 32.6 & 15s \\
    $0.1\%$ & 39.0 & 0.6s & 31.3 & 1.5s & 31.1 & 2.4s & 31.8 & 9.1s \\
    \bottomrule
  \end{tabular}
\end{table}

\begin{figure}[t!]
  \centering
   \includegraphics[width=1.0\columnwidth]{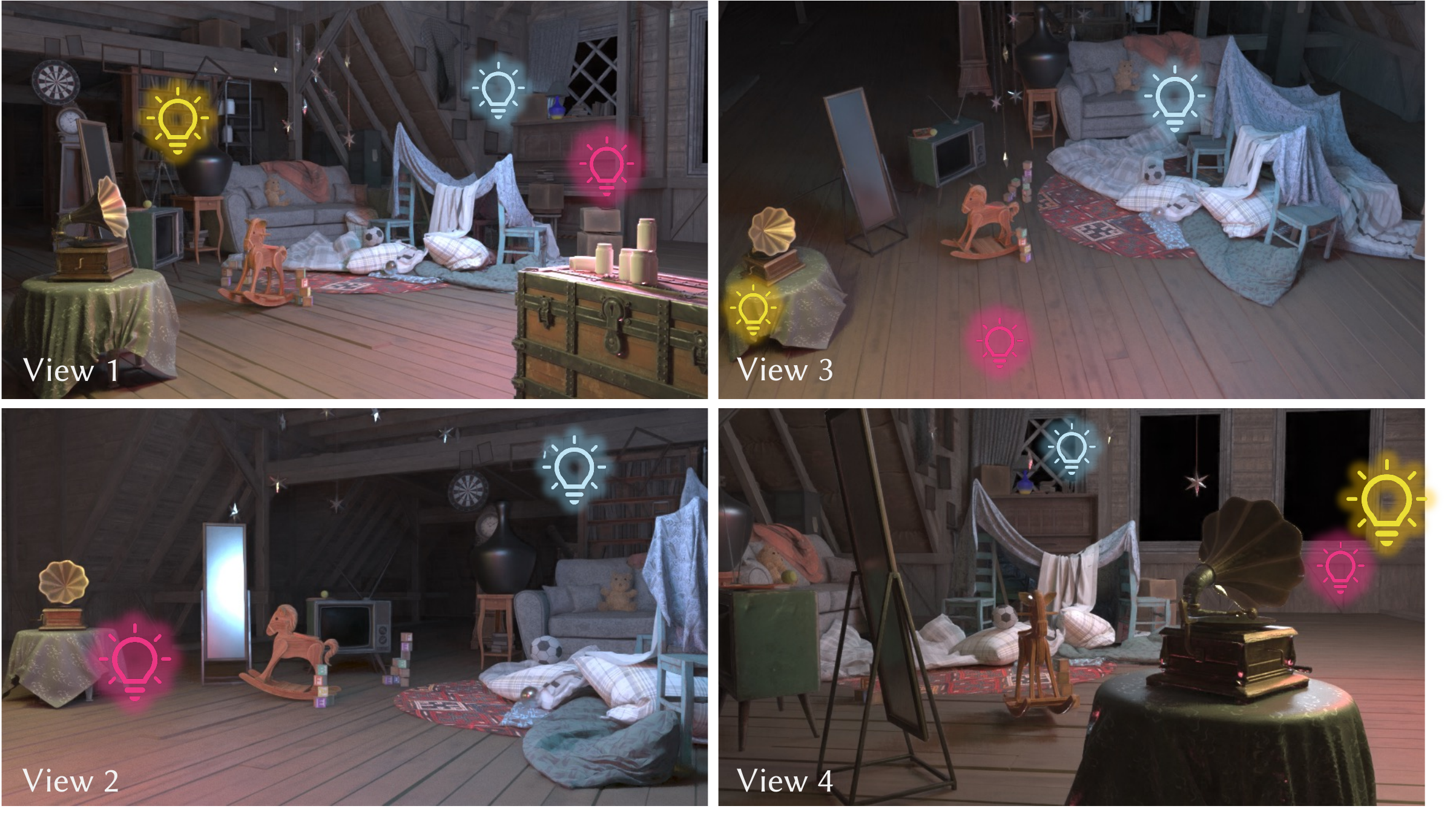}
\caption{\textbf{Multi-view lighting edits.} By training an NRP for each view, our method enables interactive and consistent lighting edits across multiple views simultaneously.}
   \label{fig:3.1-multiview}
\end{figure}

We ablate the optimization of light parameters with our method under different configurations.
Note that due to our decomposition into individual light sources (\autoref{eq:neural_image}), our method requires evaluating each light individually and the evaluation time therefore scales linearly with the number of light sources.
To support scenarios with many lights to some limited extent, we propose to evaluate the loss on a random pixel subset $\mathcal{S}$ of size $K = \lfloor \alpha \cdot H \cdot W \rfloor$, drawn uniformly without replacement at each iteration.
We ablate pixel fractions $\alpha \in \{1.0,\, 0.50,\, 0.25,\, 0.05,\, 0.01,\, 0.001\}$, ranging from the full image down to ${\sim}$462 pixels per step for a $680 \times 680$
resolution, which means a $1000\times$ reduction in per-iteration computation.
The stochastic subset provides an unbiased estimate of the full-image gradient, analogous to mini-batch SGD where pixels play the role of data samples.
We run this ablation on $N \in \{1, 3, 5, 20\}$ lights per scene with 5 optimization runs per configuration, and show quantitative results in Table~\ref{tab:cache_ablation}. Our ablation shows good robustness of the optimization at small pixel fractions, showing our method can still be performant in (modest) many light scenarios.

\begin{figure}[b!]
  \centering
   \includegraphics[width=1.0\columnwidth]{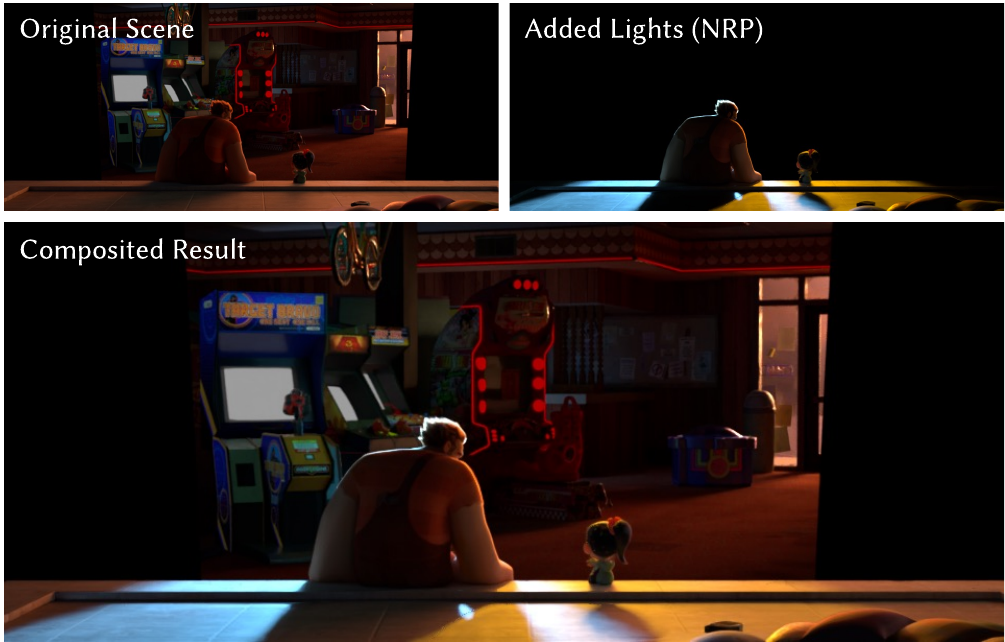}
    \caption{\textbf{Compositing Application.} Our NRP can be seamlessly combined with standard compositing workflows by training it for specific layers. In this \textit{Arcade} example, an NRP is trained on the foreground layer only, allowing to relight it while leaving the background untouched. \copyright~Disney}
   \label{fig:7_4_arcade_compositing}
\end{figure}

\begin{figure*}[h]
  \centering
   \includegraphics[width=1\linewidth]{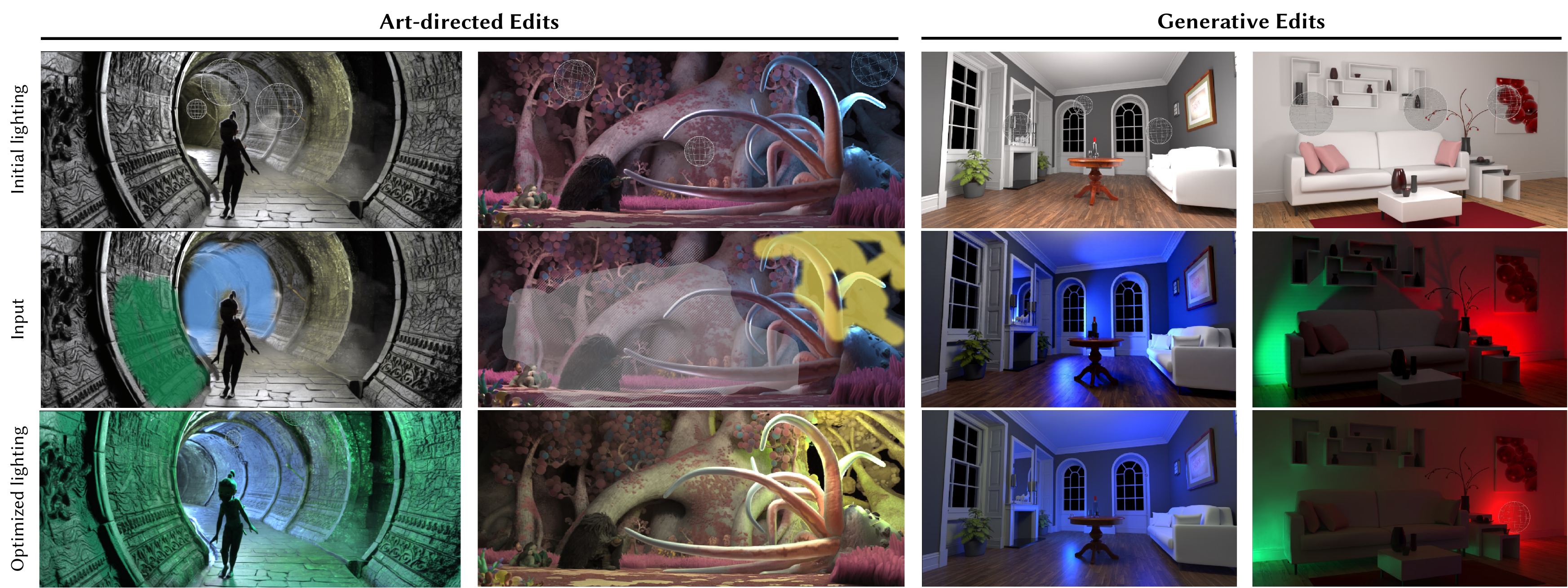}
    \caption{\textbf{Controllable relighting using our optimization pipeline.} \textbf{Left two columns (Art-directed edits):} Lighting parameters are optimized to adhere to hand-drawn scribbles depicting desired illumination phenomena. \textbf{Right two columns (Generative edits):} We leverage generative image models (Qwen-Image-Edit) to create relighting targets. In both cases, given an initial image and a target, we optimize through our NRP to extract physically plausible lighting parameters that accurately reconstruct the prompted appearance. \copyright~Disney}
   \label{fig:results_backward}
\end{figure*}

\section{Applications}

Differentiable NRPs can be leveraged for many lighting paradigms. We demonstrate three applications enabled by the compact, differentiable nature of NRPs.

\subsection{Interactive relighting}
\label{sec:interactive_relighting}
\begin{figure}[htbp]
    \centering
    \begin{subfigure}[t]{0.489\columnwidth}
        \centering
        \includegraphics[width=\linewidth]{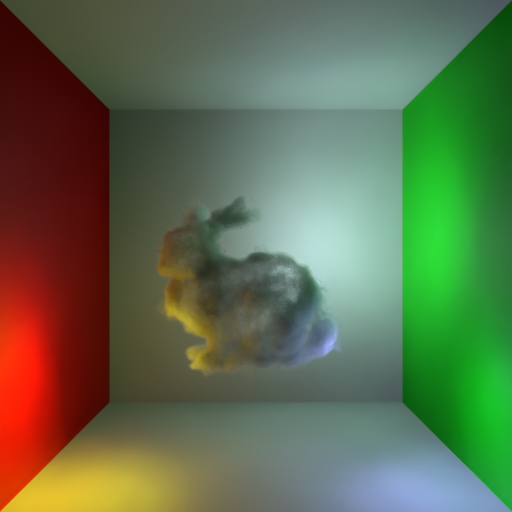}
    \end{subfigure}
    \hfill
    \begin{subfigure}[t]{0.49\columnwidth}
        \centering
        \includegraphics[width=\linewidth]{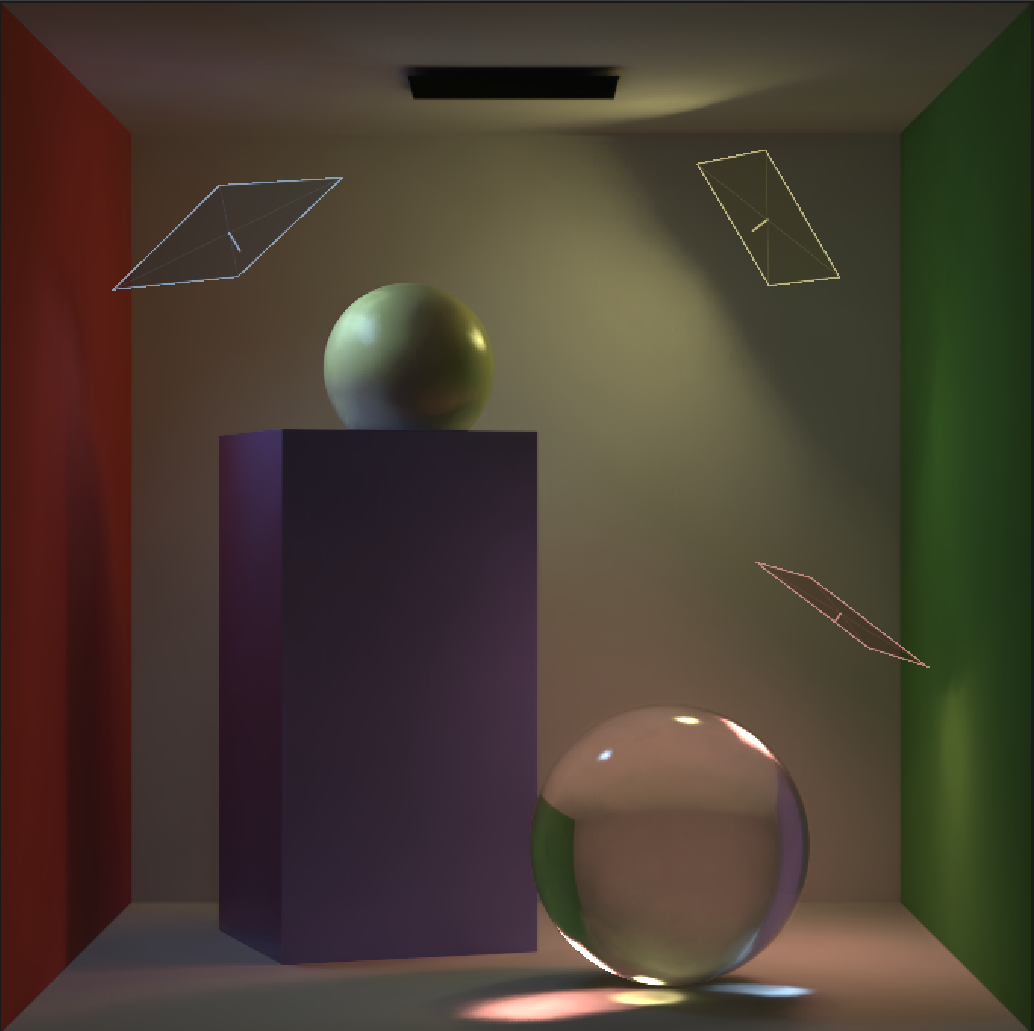}
    \end{subfigure}
    \caption{Our formulation extends to dense participating media (left) and alternative light parameterizations, such as quad lights with adjustable size, position, and orientation (right).}    \label{fig:7_1_volume_and_quadlight}
\end{figure}

First, NRP is immediately applicable in relighting applications: Given a lighting configuration, the NRP predicts the contribution of each light source to the final image pixels. As demonstrated in our experiments, the NRP network provides more accurate renderings compared to generative image-based relighting networks (\autoref{fig:comparison_diffusion}). NRPs scale gracefully to production scenes (\autoref{fig:results_production}) with complex geometry and radiometry. Our method supports relighting of volumes with spectrally non-varying extinction coefficients, and supports virtual spherical lights and virtual quad lights (Figure~\ref{fig:7_1_volume_and_quadlight}).

Second, since the NRP network is compact, unlike most neural methods, separate NRPs for different views can be loaded simultaneously into memory. This is particularly useful in production multi-view setups, where several cameras capture the same scene from different viewpoints. Traditionally, every lighting edit would need to be applied and re-rendered independently for each view. With NRPs, however, each light edit can be applied interactively and consistently across all views simultaneously (see \autoref{fig:3.1-multiview}). Similar applications for simultaneous lighting editing across entire sequences are thinkable, but have not been explored in this work.

Third, our approach can be seamlessly combined with compositing workflows; separate NRPs can be trained on separately rendered layers of a scene, allowing us to edit the lighting of individual layers without affecting other layers (Figure~\ref{fig:7_4_arcade_compositing}).

\subsection{Art-directed edits}
\label{sec:app_art_edits}
The differentiability of NRPs admits their use in optimization-based lighting edits, such as image-space shading target matching (see \autoref{fig:5_6-comparison_differentiability}) and art-directed ``light scribbling'' (see \autoref{fig:results_backward}, left). In this workflow, artists paint color scribbles directly onto the rendered image to indicate desired lighting changes, and the optimizer adjusts the underlying light parameters to match them. Constraint masks can be applied to preserve existing pixel values in regions the artist does not wish to modify, ensuring that only the targeted areas are affected. This combination of intuitive painting controls and automatic parameter recovery enables rapid iterative workflows, particularly suited for live sessions between lighting supervisors and lighting artists.

\subsection{Optimization with generative targets}
\label{sec:app_generative}
Generative image models can produce compelling lighting variations, but their outputs are not guaranteed to be physically plausible and do not provide the underlying light parameters needed for integration into a production pipeline. By using a generated image as an optimization target, our method recovers a close physically plausible lighting configuration along with its explicit parameters, bridging the gap between generative creativity and physically-based rendering. \autoref{fig:results_backward} (right) presents qualitative results demonstrating this capability, utilizing target images generated by Qwen-Image-Edit \cite{wu2025qwenimagetechnicalreport}.

\section{Limitations and Future Work}
While our model in principle generalizes to arbitrary types of light sources, their respective parameter counts directly impact the training convergence and accuracy of our neural render proxy.
We therefore leave complex light sources, such as textured lights or area lights of arbitrary shapes, as future work.

Another limitation is that our approach assumes that the light transport between any objects remains constant after rendering. Commonly used tools in the arsenal of lighters, like non-physical light-attenuation between objects or light-exclusivity for certain objects, cannot be changed after the NRP was trained. While per-layer NRPs can be used to somewhat mimic light-exclusivity, it comes at the price of training multiple NRPs.

\begin{figure}[htbp]
    \centering
    \begin{subfigure}[t]{0.489\columnwidth}
        \centering
        \includegraphics[width=\linewidth]{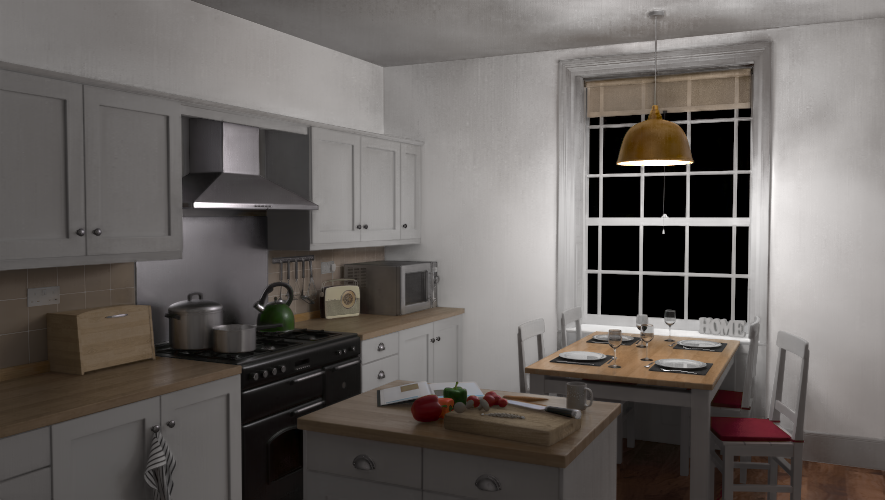}
        \caption{NRP Prediction}
    \end{subfigure}
    \hfill
    \begin{subfigure}[t]{0.49\columnwidth}
        \centering
        \includegraphics[width=\linewidth]{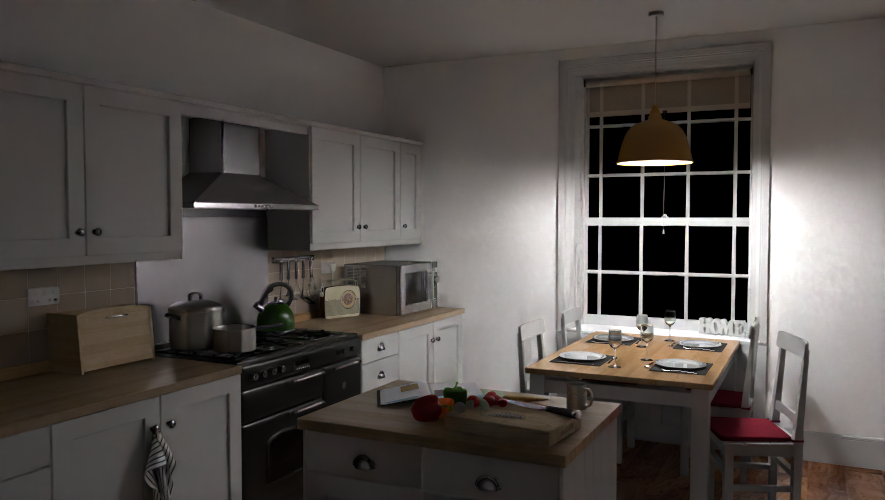}
        \caption{Ground-truth Reconstruction}
    \end{subfigure}
    \caption{\textbf{Limitations in narrow regions.} When queried in under-sampled spaces, our NRP can struggle to accurately model occlusions. As shown here, placing the light source inside the narrow cavity of the lamp results in incorrect light propagation.}
    \label{fig:7_2_limitations}
\end{figure}

Additionally, our training relies on a light-agnostic path sampling strategy. 
We currently employ material-based sampling, which produces training data with similar throughput weight to the camera. 
However, this can lead to under-sampling in occluded regions where a user might later wish to place a light source. For instance, in Figure~\ref{fig:7_2_limitations}, the region inside the lamp shade is under-sampled since no NEE was used during path sample generation, which leads to inaccurate predictions.
Investigating different sampling strategies during data generation that deliberately increase the sampling density of locations where a user might add light sources could yield significant improvements in quality.

Finally, our training pipeline is constrained by the available GPU memory, as all path-sample data is pre-loaded. 
While this enables on-the-fly generation of denoised training targets, it limits the number of samples and the quality of our training data. 
Future research could investigate out-of-core training schedules that allow on-demand loading of path data from disk during training, or partitioning of the predicted image.

\section{Conclusion}

Interactive, physically grounded lighting design on production scenes has been limited by the cost of path tracing, trapping artists in a slow cycle of parameter adjustment and re-rendering. The neural render proxy removes this bottleneck: A single light-agnostic rendering pass produces a compact, differentiable model that enables real-time forward relighting with full parametric precision, independent of scene complexity. Its differentiability further unlocks a new mode of creative control that production pipelines have traditionally lacked: Image-space objectives from art direction, paint-overs, or generative models are translated via gradient descent into the lighting parameters that produce them. Production scenes thus become interactively relightable and inversely designable through a single representation, with every edit grounded in a physical, editable, and reusable set of light configurations.

\paragraph*{Acknowledgment}
We thank Violaine Fayolle and Doriano van Essen for the lighting design on the showcased scenes, Charlotte Zhu and Daniel Teece for providing the production scenes, and Karl Li and André Mazzone for their support since the early stages of the project.
We are also grateful to Christian Döring for his help with Mitsuba 3 function freezing, and to Yotam Erel for running the relighting comparisons.

\bibliographystyle{eg-alpha-doi}
\bibliography{egbibsample}

\newcommand{\etalchar}[1]{$^{#1}$}
\begin{thebibliography}{\uppercase{LGND{\etalchar{*}}24}}

\bibitem[{\'A}fr26]{OpenImageDenoise}
\textsc{{\'A}fra A.~T.}:
\newblock {Intel\textsuperscript{\textregistered} Open Image Denoise}, 2026.
\newblock \url{https://www.openimagedenoise.org}.

\bibitem[BAC{\etalchar{*}}18]{hyperion}
\textsc{Burley B., Adler D., Chiang M. J.-Y., Driskill H., Habel R., Kelly P.,
  Kutz P., Li Y.~K., Teece D.}:
\newblock The design and evolution of disney’s hyperion renderer.
\newblock \emph{ACM Trans. Graph. 37}, 3 (July 2018).
\newblock \href {https://doi.org/10.1145/3182159} {\path{doi:10.1145/3182159}}.

\bibitem[BBJ{\etalchar{*}}21]{boss_nerd_2021}
\textsc{Boss M., Braun R., Jampani V., Barron J.~T., Liu C., Lensch H. P.~A.}:
\newblock {NeRD}: Neural reflectance decomposition from image collections.
\newblock In \emph{Proceedings of the {IEEE}/{CVF} International Conference on
  Computer Vision} (2021), pp.~12684--12694.
\newblock \href {https://doi.org/10.1109/ICCV48922.2021.01245}
  {\path{doi:10.1109/ICCV48922.2021.01245}}.

\bibitem[BFB{\etalchar{*}}25]{bharadwaj_genlit_2025}
\textsc{Bharadwaj S., Feng H., Becherini G., Fernandez~Abrevaya V., Black
  M.~J.}:
\newblock {GenLit}: Reformulating single-image relighting as video generation.
\newblock In \emph{Proceedings of the {SIGGRAPH} Asia 2025 Conference Papers}
  (2025), {SA} Conference Papers '25, Association for Computing Machinery,
  pp.~1--12.
\newblock \href {https://doi.org/10.1145/3757377.3763970}
  {\path{doi:10.1145/3757377.3763970}}.

\bibitem[BLD20]{bangaru_warped_2020}
\textsc{Bangaru S.~P., Li T.-M., Durand F.}:
\newblock Unbiased warped-area sampling for differentiable rendering.
\newblock \emph{ACM Trans. Graph. 39}, 6 (Nov. 2020).
\newblock \href {https://doi.org/10.1145/3414685.3417833}
  {\path{doi:10.1145/3414685.3417833}}.

\bibitem[BWP{\etalchar{*}}20]{bitterli_restir_2020}
\textsc{Bitterli B., Wyman C., Pharr M., Shirley P., Lefohn A., Jarosz W.}:
\newblock Spatiotemporal reservoir resampling for real-time ray tracing with
  dynamic direct lighting.
\newblock \emph{ACM Trans. Graph. 39}, 4 (Aug. 2020).
\newblock \href {https://doi.org/10.1145/3386569.3392481}
  {\path{doi:10.1145/3386569.3392481}}.

\bibitem[CA25]{careaga_physically_2025}
\textsc{Careaga C., Aksoy Y.}:
\newblock Physically controllable relighting of photographs.
\newblock In \emph{Proceedings of the Special Interest Group on Computer
  Graphics and Interactive Techniques Conference Conference Papers} (2025),
  {SIGGRAPH} Conference Papers '25, Association for Computing Machinery,
  pp.~1--10.
\newblock \href {https://doi.org/10.1145/3721238.3730666}
  {\path{doi:10.1145/3721238.3730666}}.

\bibitem[CWP{\etalchar{*}}25]{choi_scribblelight_2025}
\textsc{Choi J.~M., Wang A., Peers P., Bhattad A., Sengupta R.}:
\newblock {ScribbleLight}: Single image indoor relighting with scribbles.
\newblock In \emph{Proceedings of the {IEEE}/{CVF} Conference on Computer
  Vision and Pattern Recognition} (2025), pp.~5720--5731.
\newblock \href {https://doi.org/10.1109/CVPR52734.2025.00537}
  {\path{doi:10.1109/CVPR52734.2025.00537}}.

\bibitem[EDG{\etalchar{*}}25]{erel_practilight_2025}
\textsc{Erel Y., Dabral R., Golyanik V., Bermano A.~H., Theobalt C.}:
\newblock {PractiLight}: Practical light control using foundational diffusion
  models.
\newblock \emph{{ACM} Trans. Graph. 44}, 6 (2025), 264:1--264:11.
\newblock \href {https://doi.org/10.1145/3763342} {\path{doi:10.1145/3763342}}.

\bibitem[FR24]{fischer_zerograds_2024}
\textsc{Fischer M., Ritschel T.}:
\newblock {ZeroGrads}: Learning local surrogates for non-differentiable
  graphics.
\newblock \emph{{ACM} Transactions on Graphics 43}, 4 (2024), 1--15.
\newblock \href {https://doi.org/10.1145/3658173} {\path{doi:10.1145/3658173}}.

\bibitem[GRP22]{griffiths_outcast_2022}
\textsc{Griffiths D., Ritschel T., Philip J.}:
\newblock {OutCast}: Outdoor single-image relighting with cast shadows.
\newblock \emph{Computer Graphics Forum 41}, 2 (2022), 179--193.
\newblock \_eprint: https://onlinelibrary.wiley.com/doi/pdf/10.1111/cgf.14467.
\newblock \href {https://doi.org/10.1111/cgf.14467}
  {\path{doi:10.1111/cgf.14467}}.

\bibitem[GSY{\etalchar{*}}17]{gardner_learning_2017}
\textsc{Gardner M.-A., Sunkavalli K., Yumer E., Shen X., Gambaretto E., Gagné
  C., Lalonde J.-F.}:
\newblock Learning to predict indoor illumination from a single image.
\newblock \emph{{ACM} Trans. Graph. 36}, 6 (2017), 176:1--176:14.
\newblock \href {https://doi.org/10.1145/3130800.3130891}
  {\path{doi:10.1145/3130800.3130891}}.

\bibitem[HLN{\etalchar{*}}23]{hadadan_inverse_2023}
\textsc{Hadadan S., Lin G., Novák J., Rousselle F., Zwicker M.}:
\newblock Inverse global illumination using a neural radiometric prior.
\newblock In \emph{{ACM} {SIGGRAPH} 2023 Conference Proceedings} (2023),
  {SIGGRAPH} '23, Association for Computing Machinery, pp.~1--11.
\newblock \href {https://doi.org/10.1145/3588432.3591553}
  {\path{doi:10.1145/3588432.3591553}}.

\bibitem[HSRM25]{SiggraphCourse2025pathguiding}
\textsc{Herholz S., Sik M., Reichardt L., Manzi M.}:
\newblock Path guiding in production and recent advancements.
\newblock In \emph{Proceedings of the Special Interest Group on Computer
  Graphics and Interactive Techniques Conference Courses} (New York, NY, USA,
  2025), SIGGRAPH Courses '25, Association for Computing Machinery.
\newblock \href {https://doi.org/10.1145/3721241.3733994}
  {\path{doi:10.1145/3721241.3733994}}.

\bibitem[JLL{\etalchar{*}}24]{jin_neural_2024}
\textsc{Jin H., Li Y., Luan F., Xiangli Y., Bi S., Zhang K., Xu Z., Sun J.,
  Snavely N.}:
\newblock Neural gaffer: Relighting any object via diffusion.
\newblock In \emph{Advances in Neural Information Processing Systems} (2024),
  Globerson A., Mackey L., Belgrave D., Fan A., Paquet U., Tomczak J., Zhang
  C., (Eds.), vol.~37, Curran Associates, Inc., pp.~141129--141152.
\newblock \href {https://doi.org/10.52202/079017-4481}
  {\path{doi:10.52202/079017-4481}}.

\bibitem[JSL{\etalchar{*}}25]{jiang_differentiable_2025}
\textsc{Jiang K., Sun J.-M., Li Z., Wang D., Li T.-M., Ramamoorthi R.}:
\newblock Differentiable light transport with gaussian surfels via adapted
  radiosity for efficient relighting and geometry reconstruction.
\newblock \emph{{ACM} Trans. Graph. 44}, 6 (2025), 210:1--210:25.
\newblock \href {https://doi.org/10.1145/3763305} {\path{doi:10.1145/3763305}}.

\bibitem[JSR{\etalchar{*}}22]{jakob_mitsuba_2022}
\textsc{Jakob W., Speierer S., Roussel N., Nimier-David M., Vicini D., Zeltner
  T., Nicolet B., Crespo M., Leroy V., Zhang Z.}:
\newblock Mitsuba 3 - a retargetable forward and inverse renderer, 2022.
\newblock URL: \url{https://www.mitsuba-renderer.org/}.

\bibitem[JSRV22]{jakob_drjit_2022}
\textsc{Jakob W., Speierer S., Roussel N., Vicini D.}:
\newblock {Dr}.{Jit}: A just-in-time compiler for differentiable rendering.
\newblock \emph{{ACM} Trans. Graph. 41}, 4 (2022), 124:1--124:19.
\newblock \href {https://doi.org/10.1145/3528223.3530099}
  {\path{doi:10.1145/3528223.3530099}}.

\bibitem[KHLN17]{kutz2017spectral}
\textsc{Kutz P., Habel R., Li Y.~K., Nov\'{a}k J.}:
\newblock Spectral and decomposition tracking for rendering heterogeneous
  volumes.
\newblock \emph{ACM Trans. Graph. 36}, 4 (July 2017).
\newblock \href {https://doi.org/10.1145/3072959.3073665}
  {\path{doi:10.1145/3072959.3073665}}.

\bibitem[KP09]{kerr_evaluating_2009}
\textsc{Kerr W.~B., Pellacini F.}:
\newblock Toward evaluating lighting design interface paradigms for novice
  users.
\newblock \emph{ACM Trans. Graph. 28}, 3 (July 2009).
\newblock \href {https://doi.org/10.1145/1531326.1531332}
  {\path{doi:10.1145/1531326.1531332}}.

\bibitem[KSL05]{prt_survey_2005}
\textsc{Kautz J., Sloan P.-P., Lehtinen J.}:
\newblock Precomputed radiance transfer: theory and practice.
\newblock In \emph{ACM SIGGRAPH 2005 Courses} (New York, NY, USA, 2005),
  SIGGRAPH '05, Association for Computing Machinery, p.~1–es.
\newblock \href {https://doi.org/10.1145/1198555.1198682}
  {\path{doi:10.1145/1198555.1198682}}.

\bibitem[LADL18]{li_differentiable_2018}
\textsc{Li T.-M., Aittala M., Durand F., Lehtinen J.}:
\newblock Differentiable monte carlo ray tracing through edge sampling.
\newblock \emph{{ACM} Trans. Graph. 37}, 6 (2018), 222:1--222:11.
\newblock \href {https://doi.org/10.1145/3272127.3275109}
  {\path{doi:10.1145/3272127.3275109}}.

\bibitem[LCY{\etalchar{*}}24]{luo_intrinsicdiffusion_2024}
\textsc{Luo J., Ceylan D., Yoon J.~S., Zhao N., Philip J., Frühstück A., Li
  W., Richardt C., Wang T.}:
\newblock {IntrinsicDiffusion}: Joint intrinsic layers from latent diffusion
  models.
\newblock In \emph{{ACM} {SIGGRAPH} 2024 Conference Papers} (2024), {SIGGRAPH}
  '24, Association for Computing Machinery, pp.~1--11.
\newblock \href {https://doi.org/10.1145/3641519.3657472}
  {\path{doi:10.1145/3641519.3657472}}.

\bibitem[LDHG{\etalchar{*}}25]{lyu_intrinsicedit_2025}
\textsc{Lyu L., Deschaintre V., Hold-Geoffroy Y., Hašan M., Yoon J.~S.,
  Leimkühler T., Theobalt C., Georgiev I.}:
\newblock {IntrinsicEdit}: Precise generative image manipulation in intrinsic
  space.
\newblock \emph{{ACM} Trans. Graph. 44}, 4 (2025), 106:1--106:13.
\newblock \href {https://doi.org/10.1145/3731173} {\path{doi:10.1145/3731173}}.

\bibitem[LGND{\etalchar{*}}24]{liang_photorealistic_2025}
\textsc{Liang R., Gojcic Z., Nimier-David M., Acuna D., Vijaykumar N., Fidler
  S., Wang Z.}:
\newblock Photorealistic object insertion with diffusion-guided inverse
  rendering.
\newblock In \emph{Computer Vision – ECCV 2024: 18th European Conference,
  Milan, Italy, September 29–October 4, 2024, Proceedings, Part LXI} (Berlin,
  Heidelberg, 2024), Springer-Verlag, p.~446–465.
\newblock \href {https://doi.org/10.1007/978-3-031-73030-6_25}
  {\path{doi:10.1007/978-3-031-73030-6_25}}.

\bibitem[LHEN{\etalchar{*}}24]{lipp_view-independent_2024}
\textsc{Lipp L., Hahn D., Ecormier-Nocca P., Rist F., Wimmer M.}:
\newblock View-independent adjoint light tracing for lighting design
  optimization.
\newblock \emph{{ACM} Trans. Graph. 43}, 3 (2024), 35:1--35:16.
\newblock \href {https://doi.org/10.1145/3662180} {\path{doi:10.1145/3662180}}.

\bibitem[LHG{\etalchar{*}}25]{liang_luxdit_2025}
\textsc{Liang R., He K., Gojcic Z., Gilitschenski I., Fidler S., Vijaykumar N.,
  Wang Z.}:
\newblock {LuxDiT}: Lighting estimation with video diffusion transformer, 2025.
\newblock \href {http://arxiv.org/abs/2509.03680 [cs]} {\path{arXiv:2509.03680
  [cs]}}, \href {https://doi.org/10.48550/arXiv.2509.03680}
  {\path{doi:10.48550/arXiv.2509.03680}}.

\bibitem[MESK22]{muller_instant_2022}
\textsc{Müller T., Evans A., Schied C., Keller A.}:
\newblock Instant neural graphics primitives with a multiresolution hash
  encoding.
\newblock \emph{{ACM} Trans. Graph. 41}, 4 (2022), 102:1--102:15.
\newblock \href {https://doi.org/10.1145/3528223.3530127}
  {\path{doi:10.1145/3528223.3530127}}.

\bibitem[MGN17]{muller_pathguiding_2017}
\textsc{M\"{u}ller T., Gross M., Nov\'{a}k J.}:
\newblock Practical path guiding for efficient light-transport simulation.
\newblock \emph{Comput. Graph. Forum 36}, 4 (July 2017), 91--100.
\newblock \href {https://doi.org/10.1111/cgf.13227}
  {\path{doi:10.1111/cgf.13227}}.

\bibitem[MHT{\etalchar{*}}25]{magar_lightlab_2025}
\textsc{Magar N., Hertz A., Tabellion E., Pritch Y., Rav-Acha A., Shamir A.,
  Hoshen Y.}:
\newblock {LightLab}: Controlling light sources in images with diffusion
  models.
\newblock In \emph{{SIGGRAPH} Conference Papers '25: Special Interest Group on
  Computer Graphics and Interactive Techniques Conference Conference Papers}
  (2025), {ACM}, pp.~1--11.
\newblock \href {https://doi.org/10.1145/3721238.3730696}
  {\path{doi:10.1145/3721238.3730696}}.

\bibitem[MRNK21]{muller_nrc_2021}
\textsc{M\"{u}ller T., Rousselle F., Nov\'{a}k J., Keller A.}:
\newblock Real-time neural radiance caching for path tracing.
\newblock \emph{ACM Trans. Graph. 40}, 4 (July 2021).
\newblock \href {https://doi.org/10.1145/3450626.3459812}
  {\path{doi:10.1145/3450626.3459812}}.

\bibitem[M{\"u}l21]{tiny-cuda-nn}
\textsc{M{\"u}ller T.}:
\newblock {tiny-cuda-nn}, 4 2021.
\newblock URL: \url{https://github.com/NVlabs/tiny-cuda-nn}.

\bibitem[NDSRJ20]{nimier-david_radiative_2020}
\textsc{Nimier-David M., Speierer S., Ruiz B., Jakob W.}:
\newblock Radiative backpropagation: an adjoint method for lightning-fast
  differentiable rendering.
\newblock \emph{{ACM} Trans. Graph. 39}, 4 (2020), 146:146:1--146:146:15.
\newblock \href {https://doi.org/10.1145/3386569.3392406}
  {\path{doi:10.1145/3386569.3392406}}.

\bibitem[PBMF07]{pellacini_lighting_2007}
\textsc{Pellacini F., Battaglia F., Morley R.~K., Finkelstein A.}:
\newblock Lighting with paint.
\newblock \emph{ACM Trans. Graph. 26}, 2 (June 2007), 9–es.
\newblock \href {https://doi.org/10.1145/1243980.1243983}
  {\path{doi:10.1145/1243980.1243983}}.

\bibitem[PCS{\etalchar{*}}24]{phongthawee_diffusionlight_2024}
\textsc{Phongthawee P., Chinchuthakun W., Sinsunthithet N., Jampani V., Raj A.,
  Khungurn P., Suwajanakorn S.}:
\newblock {DiffusionLight}: Light probes for free by painting a chrome ball.
\newblock In \emph{Proceedings of the {IEEE}/{CVF} Conference on Computer
  Vision and Pattern Recognition} (2024), pp.~98--108.
\newblock \href {https://doi.org/10.1109/CVPR52733.2024.00018}
  {\path{doi:10.1109/CVPR52733.2024.00018}}.

\bibitem[Pha18]{TOG-prod-rendering}
\textsc{Pharr M.}:
\newblock Guest editor’s introduction: Special issue on production rendering.
\newblock \emph{ACM Trans. Graph. 37}, 3 (July 2018).
\newblock \href {https://doi.org/10.1145/3212511} {\path{doi:10.1145/3212511}}.

\bibitem[PTG02]{pellacini_shadow_2002}
\textsc{Pellacini F., Tole P., Greenberg D.~P.}:
\newblock A user interface for interactive cinematic shadow design.
\newblock \emph{ACM Trans. Graph. 21}, 3 (July 2002), 563–566.
\newblock \href {https://doi.org/10.1145/566654.566617}
  {\path{doi:10.1145/566654.566617}}.

\bibitem[PVL{\etalchar{*}}05]{pellacini_lpics_2005}
\textsc{Pellacini F., Vidim\v{c}e K., Lefohn A., Mohr A., Leone M., Warren J.}:
\newblock Lpics: a hybrid hardware-accelerated relighting engine for computer
  cinematography.
\newblock \emph{ACM Trans. Graph. 24}, 3 (July 2005), 464–470.
\newblock \href {https://doi.org/10.1145/1073204.1073214}
  {\path{doi:10.1145/1073204.1073214}}.

\bibitem[RBRD22]{neural_prt_2022}
\textsc{Rainer G., Bousseau A., Ritschel T., Drettakis G.}:
\newblock Neural precomputed radiance transfer.
\newblock \emph{Computer Graphics Forum (Proceedings of the Eurographics
  conference) 41}, 2 (April 2022).
\newblock \href {https://doi.org/10.1111/cgf.14478}
  {\path{doi:10.1111/cgf.14478}}.

\bibitem[RDL{\etalchar{*}}15]{ren_image_2015}
\textsc{Ren P., Dong Y., Lin S., Tong X., Guo B.}:
\newblock Image based relighting using neural networks.
\newblock \emph{{ACM} Trans. Graph. 34}, 4 (2015), 111:1--111:12.
\newblock \href {https://doi.org/10.1145/2766899} {\path{doi:10.1145/2766899}}.

\bibitem[RHP{\etalchar{*}}24]{ren_lightformer_2024}
\textsc{Ren H., Huo Y., Peng Y., Sheng H., Xue W., Huang H., Lan J., Wang R.,
  Bao H.}:
\newblock Lightformer: Light-oriented global neural rendering in dynamic scene.
\newblock \emph{ACM Trans. Graph. 43}, 4 (July 2024).
\newblock \href {https://doi.org/10.1145/3658229} {\path{doi:10.1145/3658229}}.

\bibitem[RKKS{\etalchar{*}}07]{ragankelley_lightspeed_2007}
\textsc{Ragan-Kelley J., Kilpatrick C., Smith B.~W., Epps D., Green P., Hery
  C., Durand F.}:
\newblock The lightspeed automatic interactive lighting preview system.
\newblock \emph{ACM Trans. Graph. 26}, 3 (July 2007), 25–es.
\newblock \href {https://doi.org/10.1145/1276377.1276409}
  {\path{doi:10.1145/1276377.1276409}}.

\bibitem[RXL{\etalchar{*}}23]{raghavan_neural_2023}
\textsc{Raghavan N., Xiao Y., Lin K.-E., Sun T., Bi S., Xu Z., Li T.-M.,
  Ramamoorthi R.}:
\newblock Neural free-viewpoint relighting for glossy indirect illumination.
\newblock \emph{Computer Graphics Forum (Proc. EGSR 2023) 42}, 4 (2023).

\bibitem[SBD{\etalchar{*}}25]{shen_stglight_2025}
\textsc{Shen S., Bao Z., Ding H., Xu W., Lai T., Xiao C.}:
\newblock {STGlight}: Online indoor lighting estimation via spatio-temporal
  gaussian fusion.
\newblock \emph{{ACM} Trans. Graph. 44}, 6 (2025), 176:1--176:14.
\newblock \href {https://doi.org/10.1145/3763350} {\path{doi:10.1145/3763350}}.

\bibitem[SDS{\etalchar{*}}93]{schoeneman_painting_1993}
\textsc{Schoeneman C., Dorsey J., Smits B., Arvo J., Greenberg D.}:
\newblock Painting with light.
\newblock In \emph{Proceedings of the 20th Annual Conference on Computer
  Graphics and Interactive Techniques} (New York, NY, USA, 1993), SIGGRAPH '93,
  Association for Computing Machinery, p.~143–146.
\newblock \href {https://doi.org/10.1145/166117.166135}
  {\path{doi:10.1145/166117.166135}}.

\bibitem[TKC19]{tillet2019triton}
\textsc{Tillet P., Kung H.~T., Cox D.}:
\newblock Triton: an intermediate language and compiler for tiled neural
  network computations.
\newblock In \emph{Proceedings of the 3rd ACM SIGPLAN International Workshop on
  Machine Learning and Programming Languages} (New York, NY, USA, 2019), MAPL
  2019, Association for Computing Machinery, pp.~10--19.
\newblock \href {https://doi.org/10.1145/3315508.3329973}
  {\path{doi:10.1145/3315508.3329973}}.

\bibitem[TWZ25]{tong_spatiotemporally_2025}
\textsc{Tong M., Wu R., Zheng C.}:
\newblock Spatiotemporally consistent indoor lighting estimation with diffusion
  priors.
\newblock In \emph{Proceedings of the Special Interest Group on Computer
  Graphics and Interactive Techniques Conference Conference Papers} (2025),
  {SIGGRAPH} Conference Papers '25, Association for Computing Machinery,
  pp.~1--11.
\newblock \href {https://doi.org/10.1145/3721238.3730749}
  {\path{doi:10.1145/3721238.3730749}}.

\bibitem[VSJ21]{vicini_path_2021}
\textsc{Vicini D., Speierer S., Jakob W.}:
\newblock Path replay backpropagation: differentiating light paths using
  constant memory and linear time.
\newblock \emph{{ACM} Trans. Graph. 40}, 4 (2021), 108:1--108:14.
\newblock \href {https://doi.org/10.1145/3450626.3459804}
  {\path{doi:10.1145/3450626.3459804}}.

\bibitem[WFR25]{wang_stochastic_2025}
\textsc{Wang Z., Fischer M., Ritschel T.}:
\newblock Stochastic gradient estimation for higher-order differentiable
  rendering.
\newblock In \emph{Proceedings of the {IEEE}/{CVF} International Conference on
  Computer Vision} (2025), pp.~28198--28206.
\newblock \href {https://doi.org/10.1109/ICCV51701.2025.02618}
  {\path{doi:10.1109/ICCV51701.2025.02618}}.

\bibitem[WJZ{\etalchar{*}}25]{wang_humanir-mgi_2025}
\textsc{Wang R., Jiang Y., Zhu H., Luo F., Xiao C.}:
\newblock {HumanIR}-{MGI}: human inverse rendering via jointly optimizing
  geometry, material, and illumination.
\newblock \emph{The Visual Computer 41}, 9 (2025), 6969--6982.
\newblock \href {https://doi.org/10.1007/s00371-025-04035-z}
  {\path{doi:10.1007/s00371-025-04035-z}}.

\bibitem[WLZ{\etalchar{*}}25]{wu2025qwenimagetechnicalreport}
\textsc{Wu C., Li J., Zhou J., Lin J., Gao K., Yan K., ming Yin S., Bai S., Xu
  X., Chen Y., Chen Y., Tang Z., Zhang Z., Wang Z., Yang A., Yu B., Cheng C.,
  Liu D., Li D., Zhang H., Meng H., Wei H., Ni J., Chen K., Cao K., Peng L., Qu
  L., Wu M., Wang P., Yu S., Wen T., Feng W., Xu X., Wang Y., Zhang Y., Zhu Y.,
  Wu Y., Cai Y., Liu Z.}:
\newblock Qwen-image technical report, 2025.
\newblock URL: \url{https://arxiv.org/abs/2508.02324}, \href
  {http://arxiv.org/abs/2508.02324} {\path{arXiv:2508.02324}}.

\bibitem[WMB{\etalchar{*}}25]{wu_wosc_2025}
\textsc{Wu L., Morrical N., Bangaru S.~P., Sawhney R., Zhao S., Wyman C.,
  Ramamoorthi R., Lefohn A.}:
\newblock Unbiased differential visibility using fixed-step
  walk-on-spherical-caps and closest silhouettes.
\newblock \emph{ACM Trans. Graph. 44}, 4 (July 2025).
\newblock \href {https://doi.org/10.1145/3731174} {\path{doi:10.1145/3731174}}.

\bibitem[XGK{\etalchar{*}}25]{xing_luminet_2025}
\textsc{Xing X., Groh K., Karaoglu S., Gevers T., Bhattad A.}:
\newblock {LumiNet}: Latent intrinsics meets diffusion models for indoor scene
  relighting.
\newblock In \emph{Proceedings of the {IEEE}/{CVF} Conference on Computer
  Vision and Pattern Recognition} (2025), pp.~442--452.
\newblock \href {https://doi.org/10.1109/CVPR52734.2025.00050}
  {\path{doi:10.1109/CVPR52734.2025.00050}}.

\bibitem[YPD{\etalchar{*}}24]{yan_variance_2024}
\textsc{Yan K., Pegoraro V., Droske M., Vorba J., Zhao S.}:
\newblock Differentiating variance for variance-aware inverse rendering.
\newblock In \emph{SIGGRAPH Asia 2024 Conference Papers} (New York, NY, USA,
  2024), SA '24, Association for Computing Machinery.
\newblock \href {https://doi.org/10.1145/3680528.3687603}
  {\path{doi:10.1145/3680528.3687603}}.

\bibitem[ZDG{\etalchar{*}}24]{zeng_rgbx_2024}
\textsc{Zeng Z., Deschaintre V., Georgiev I., Hold-Geoffroy Y., Hu Y., Luan F.,
  Yan L.-Q., Hašan M.}:
\newblock {RGB}<->x: Image decomposition and synthesis using material- and
  lighting-aware diffusion models.
\newblock In \emph{{ACM} {SIGGRAPH} 2024 Conference Papers} (2024), {SIGGRAPH}
  '24, Association for Computing Machinery, pp.~1--11.
\newblock \href {https://doi.org/10.1145/3641519.3657445}
  {\path{doi:10.1145/3641519.3657445}}.

\bibitem[ZDP{\etalchar{*}}24]{zeng_dilightnet_2024}
\textsc{Zeng C., Dong Y., Peers P., Kong Y., Wu H., Tong X.}:
\newblock {DiLightNet}: Fine-grained lighting control for diffusion-based image
  generation.
\newblock In \emph{{SIGGRAPH} '24: Special Interest Group on Computer Graphics
  and Interactive Techniques Conference} (2024), {ACM}, pp.~1--12.
\newblock \href {https://doi.org/10.1145/3641519.3657396}
  {\path{doi:10.1145/3641519.3657396}}.

\bibitem[ZDP{\etalchar{*}}25]{zeng_renderformer_2025}
\textsc{Zeng C., Dong Y., Peers P., Wu H., Tong X.}:
\newblock Renderformer: Transformer-based neural rendering of triangle meshes
  with global illumination.
\newblock In \emph{Proceedings of the Special Interest Group on Computer
  Graphics and Interactive Techniques Conference Conference Papers} (New York,
  NY, USA, 2025), SIGGRAPH Conference Papers '25, Association for Computing
  Machinery.
\newblock \href {https://doi.org/10.1145/3721238.3730595}
  {\path{doi:10.1145/3721238.3730595}}.

\bibitem[ZFT{\etalchar{*}}21]{zhang_neural_2021}
\textsc{Zhang X., Fanello S., Tsai Y.-T., Sun T., Xue T., Pandey R.,
  Orts-Escolano S., Davidson P., Rhemann C., Debevec P., Barron J.~T.,
  Ramamoorthi R., Freeman W.~T.}:
\newblock Neural light transport for relighting and view synthesis.
\newblock \emph{{ACM} Trans. Graph. 40}, 1 (2021), 9:1--9:17.
\newblock \href {https://doi.org/10.1145/3446328} {\path{doi:10.1145/3446328}}.

\bibitem[ZGF{\etalchar{*}}25]{zhang_unilight_2025}
\textsc{Zhang Z., Georgiev I., Fischer M., Hold-Geoffroy Y., Lalonde J.-F.,
  Deschaintre V.}:
\newblock {UniLight}: A unified representation for lighting, 2025.
\newblock \href {http://arxiv.org/abs/2512.04267 [cs]} {\path{arXiv:2512.04267
  [cs]}}, \href {https://doi.org/10.48550/arXiv.2512.04267}
  {\path{doi:10.48550/arXiv.2512.04267}}.

\bibitem[ZLW{\etalchar{*}}21]{zhang_physg_2021}
\textsc{Zhang K., Luan F., Wang Q., Bala K., Snavely N.}:
\newblock {PhySG}: Inverse rendering with spherical gaussians for physics-based
  material editing and relighting.
\newblock In \emph{Proceedings of the {IEEE}/{CVF} Conference on Computer
  Vision and Pattern Recognition} (2021), pp.~5453--5462.
\newblock \href {https://doi.org/10.1109/CVPR46437.2021.00541}
  {\path{doi:10.1109/CVPR46437.2021.00541}}.

\bibitem[ZMY{\etalchar{*}}20]{zhang_pathspace_2020}
\textsc{Zhang C., Miller B., Yan K., Gkioulekas I., Zhao S.}:
\newblock Path-space differentiable rendering.
\newblock \emph{ACM Trans. Graph. 39}, 4 (Aug. 2020).
\newblock \href {https://doi.org/10.1145/3386569.3392383}
  {\path{doi:10.1145/3386569.3392383}}.

\bibitem[ZRA25]{zhang_scaling_2025}
\textsc{Zhang L., Rao A., Agrawala M.}:
\newblock Scaling in-the-wild training for diffusion-based illumination
  harmonization and editing by imposing consistent light transport.
\newblock In \emph{International Conference on Learning Representations}
  (2025).
\newblock URL:
  \url{https://proceedings.iclr.cc/paper_files/paper/2025/hash/1cce374fec7829c24a97ae0e91cd0999-Abstract-Conference.html}.

\end{thebibliography}

\includepdf[pages=-]{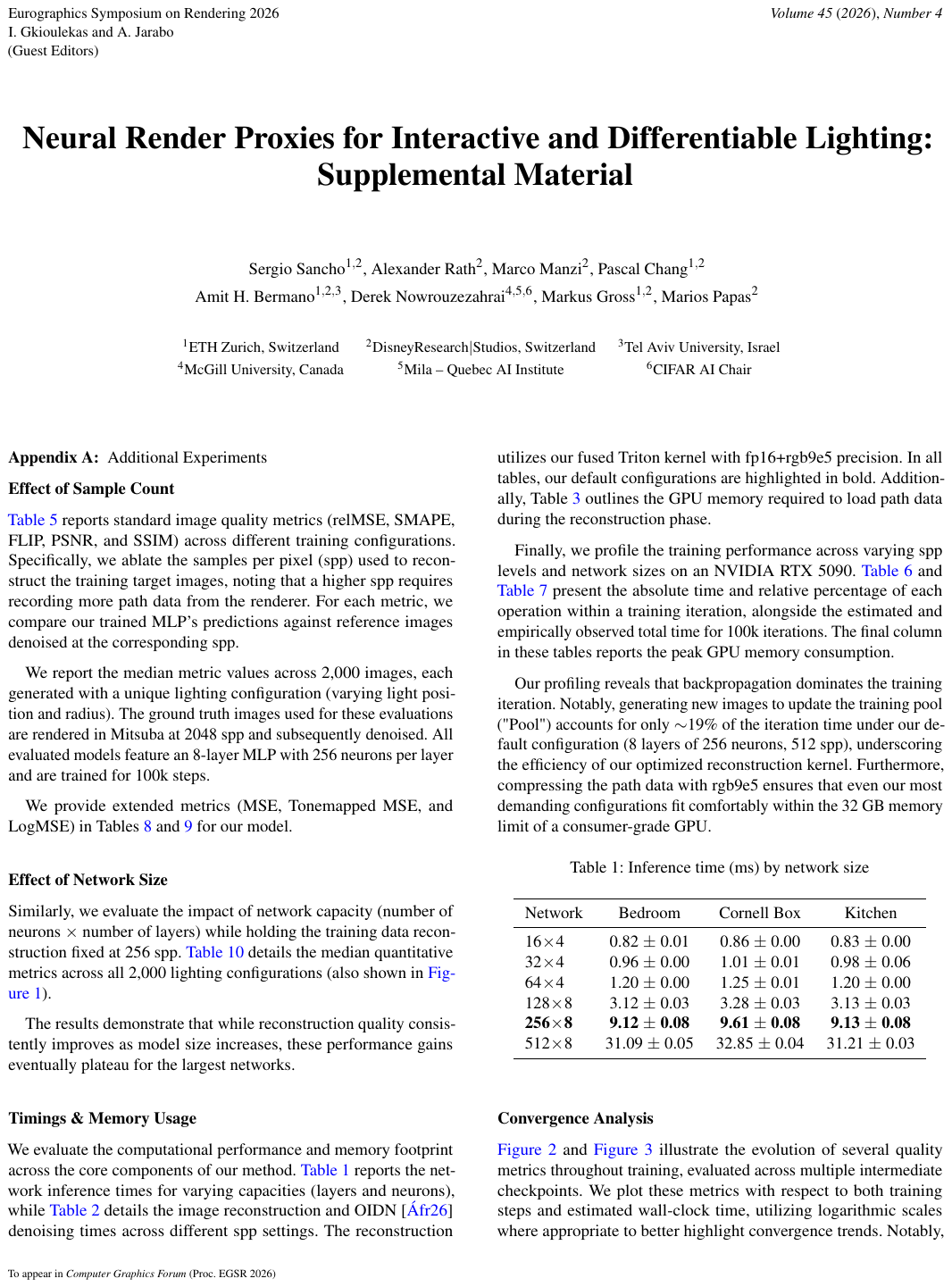}

\end{document}